\documentclass[preprint]{aastex}

\slugcomment{}

\shortauthors{Matheson et al.}
\shorttitle{Supernova 1993J}

\begin{document}


\title{Detailed Analysis of Early to Late-Time Spectra of
Supernova 1993J}

\author{Thomas Matheson\altaffilmark{1}, Alexei
V. Filippenko\altaffilmark{1}, Luis C. Ho\altaffilmark{2}, Aaron
J. Barth\altaffilmark{3}, and Douglas C. Leonard\altaffilmark{1}}
\altaffiltext{1}{Department of Astronomy, University of California,
    Berkeley, CA 94720-3411; matheson, alex, leonard@astron.berkeley.edu}
\altaffiltext{2}{The Observatories of the Carnegie Institution of
    Washington, 813 Santa Barbara Street, Pasadena, CA 91101-1292;
    lho@ociw.edu}
\altaffiltext{3}{Harvard-Smithsonian Center for Astrophysics,
    60 Garden Street, Cambridge, MA 02138; abarth@cfa.harvard.edu}



\begin{abstract}

We present a detailed study of line structure in early to late-time
spectra of Supernova (SN) 1993J.  Spectra during the nebular phase,
but within the first two years after explosion, exhibit small-scale
structure in the emission lines of some species, notably oxygen and
magnesium, showing that the ejecta of SN 1993J are clumpy.  On the
other hand, a lack of structure in emission lines of calcium implies
that the source of calcium emission is uniformly distributed
throughout the ejecta.  These results are interpreted as evidence that
oxygen emission originates in clumpy, newly synthesized material,
while calcium emission arises from material pre-existing in the
atmosphere of the progenitor.  Spectra spanning the range $433-2454$
days after the explosion show box-like profiles for the emission
lines, clearly indicating circumstellar interaction in a roughly
spherical shell.  This is interpreted within the Chevalier \& Fransson
(1994) model for SNe interacting with mass lost during prior stellar
winds.  At very late times, the emission lines have a two-horned
profile, implying the formation of a somewhat flattened or disk-like
structure that is a significant source of emission.  The very high
signal-to-noise ratio spectra are used to demonstrate the potential
significance of misinterpretation of telluric absorption lines in the
spectra of bright SNe.
\end{abstract}



\section{Introduction}

Supernova 1993J presented a great opportunity in the study of
supernovae (SNe).  It occurred in the nearby galaxy M81 (NGC 3031; $d$
= 3.6 Mpc; Freedman et al. 1994) and reached a maximum brightness of
$m_V = 10.8$ mag (e.g., Richmond et al. 1996).  This allowed extremely
detailed observations to be taken over a long time interval.  The fact
that SN 1993J underwent a transformation from a Type II to a Type IIb,
strengthening the link between SNe II and SNe Ib as core-collapse
events, only enhanced the importance of this object (see, e.g.,
Wheeler \& Filippenko 1996; Matheson et al. 2000, hereinafter Paper I,
for detailed discussions of the nature of SN 1993J).

We have previously presented analyses of the spectra of SN 1993J
obtained during its first year (Filippenko, Matheson, \& Ho 1993,
hereinafter FMH93; Filippenko, Matheson, \& Barth 1994, hereinafter
FMB94), while Paper I shows our complete low-resolution spectroscopic
data set.  The early spectra and the transition from SN II to SN IIb
are covered by FMH93, while the onset of the nebular phase and the
beginning of signs of circumstellar interaction are discussed by
FMB94.  In this paper, we take a closer look at the details of some of
the earlier spectra and we present the analysis of late-time spectra
of SN 1993J.  The clumpy nature of SN 1993J, as revealed in the
details of spectra from early to nebular times, is presented in \S 2.
The late-time spectra and their evidence of circumstellar interaction
are discussed in \S 3; we summarize our conclusions in \S 4.  In the
Appendix, we discuss the potential impact of telluric absorption on
the interpretation of the spectra of bright SNe and the implications
this has for future SN studies.

All of the low-dispersion spectra discussed herein were obtained at
the Lick and Keck Observatories.  In addition, a single
high-dispersion spectrum was obtained using the Keck~I 10-m telescope.
The details of the observations, including instrumental configuration,
dates of observation, and observational parameters, are presented in
Paper I.  We follow Lewis et al. (1994) in adopting 1993 March 27.5 UT
(JD 2,449,074) as the date of explosion.

\section{Clumps}

\subsection{Clumps in Other Supernovae}

The mottled appearance of supernova remnants (SNRs) has often been
interpreted as evidence for clumps in the ejecta of SNe.  For example,
fast-moving features have been identified in Cas A for quite some time
(e.g., Baade \& Minkowski 1954), and modern studies continue to
interpret the structure of Cas A as the result of clumpy SN ejecta
(e.g., Anderson et al. 1994).  The discovery of metal-enriched
fragments outside the boundary of the Vela SNR (Aschenbach et
al. 1995), a high-velocity ``bullet'' in Vela (Strom et al. 1995), and
fast-moving oxygen filaments in Puppis A (and other remnants; Winkler
\& Kirshner 1985, and references therein) all support the contention
that SN ejecta are clumped.  There is also a theoretical underpinning
for the expectation of clumps in the ejecta of core-collapse SNe.
This includes the necessity of mixing in SNe Ib/c (e.g., Wheeler et
al. 1987; Shigeyama et al. 1990; Hachisu et al. 1991) and the presence
of instabilities, such as convective instabilities in neutrino-driven
explosion mechanisms and Rayleigh-Taylor instabilities during the
expansion of the ejecta (e.g., Kifonidis et al. 2000, and references
therein).  In fact, with a small core mass ($3-4 M_{\sun}$), SN 1993J
should have relatively large Rayleigh-Taylor instabilities that mix
the core (Iwamoto et al. 1997).

The observation of small-scale structure in SN line emission requires
a spectrum with a high signal-to-noise (S/N) ratio and the spectral
resolution to distinguish narrow features.  This in turn necessitates
a bright source and an efficient detector with at least moderate
dispersion ($\sim 1-10$ \AA/pixel).  These two requirements for SNe
were first fulfilled with SN 1985F.  Analysis of spectra of the Type
Ib SN 1985F revealed several distinct clumps in the [\ion{O}{1}]
$\lambda\lambda$6300, 6364 lines as well (Filippenko \& Sargent 1989).
These clumps had widths $\sim 100-250$ km~s$^{-1}$ with amplitudes of
$2-10$\%.  The small-scale features changed slightly over time,
becoming narrower and relatively stronger.  The \ion{Mg}{1}]
$\lambda$4571 also had features, but they did not match those of the
oxygen lines.

The bright, nearby SN 1987A provided another opportunity to study SN
spectra in detail.  While examining the [\ion{O}{1}]
$\lambda\lambda$6300, 6364 doublet of SN 1987A, Stathakis et
al. (1991) discovered what appeared to be a sawtooth profile on the
tops of the lines: there were $\sim$ 50 small-scale features with a
typical size of 80 km~s$^{-1}$ (full width at half maximum [FWHM]) and
amplitudes of $3-10$\% of the global line profile.  The structures
persisted over time and were consistent between the two doublet lines.
Clumps were also observed in the H$\alpha$ line (Hanuschik et
al. 1993).  These had widths of $400-500$ km~s$^{-1}$ with amplitudes
of $7-10$\% of the overall line strength.  The H$\alpha$ clumps were
present over several hundred days, but individual clumps were not
persistent.  Hanuschik et al. (1993) interpreted this as changes in
the level of dust extinction along the lines of sight to the
individual clumps.  Spyromilio \& Pinto (1991) found that the oxygen
density derived from the line intensity ratios of [\ion{O}{1}]
$\lambda$6300 and [\ion{O}{1}] $\lambda$6364 gave too large an oxygen
mass if the oxygen filled the entire volume suggested by the expansion
velocity; a low filling factor for oxygen was consistent with the
calculated density, volume at that epoch, and a reasonable oxygen
mass.  Such a low filling factor can be interpreted as clumpiness.
Further evidence for clumps in the ejecta of SN 1987A is presented by
Spyromilio, Stathakis, \& Meurer (1993), who found a common feature in
emission lines of H$\alpha$, [\ion{Fe}{2}] $\lambda$7155, and
[\ion{Ca}{2}] $\lambda\lambda$7291, 7324.  Clumps were also seen at
later times in infrared (IR) emission lines of iron in SN 1987A
(Spyromilio, Meikle, \& Allen 1990; Haas et al. 1990).  The pattern
seen in the oxygen lines was not replicated in these other species.

The Type II SN 1988A may also have
had clumps.  They were not apparent in the spectra, but following the
analysis of the [\ion{O}{1}] $\lambda\lambda$6300, 6364 doublet in SN
1987A of Spyromilio \& Pinto (1991), Spyromilio (1991) derived a low
filling factor for oxygen in SN 1988A of 0.05, and thus deduced a
clumpy nature for the source of the oxygen emission.  In addition,
Fassia et al. (1998) invoked clumps of helium to explain the strength
of \ion{He}{1} $\lambda$10830 line in the Type II SN 1995V given their
model of $^{56}$Ni mixing, although the clumps were not evident from
the spectra.

SN 1993J, as noted before, was extremely bright, so any clumps in its
spectrum should be readily detectable.  Li et al. (1994) reported
clumps in their spectra of SN 1993J.  Wang \& Hu (1994) noted the
presence of narrow components in the forbidden lines of oxygen, and
they used the clumpy nature of the SN to explain apparent blueshifts
in the emission lines of [\ion{O}{1}] $\lambda$5577 and [\ion{O}{1}]
$\lambda$6300.  Spyromilio (1994, hereinafter S94) presented a much
more detailed analysis of the clumps visible in the spectra of SN
1993J and also concluded that there is large-scale anisotropy in the
SN based on the blueshifts of the lines (oxygen and magnesium) as well
as the differences in the distribution of clumps both between and
within the observed atomic species.  This interpretation of the
blueshifts was questioned by Houck \& Fransson (1995), however.  They
concluded that the blueshifts are only apparent, and did not represent
physical asymmetries in the lines.  The [\ion{O}{1}]
$\lambda\lambda$6300, 6364 doublet can be dramatically affected by
scattering of H$\alpha$ if oxygen and hydrogen are distributed in
shells, and this would mimic the observed asymmetry.  Blending affects
the other lines, with [\ion{O}{1}] $\lambda$5577 contaminated by
[\ion{Fe}{2}] $\lambda$5536 and [\ion{Co}{2}] $\lambda$5526.  The
\ion{Mg}{1}] $\lambda$4571 line is affected by \ion{Fe}{2} multiplets
near 4600 \AA.

\subsection{Analysis of the Clumps}

As the brightest supernova in the northern hemisphere since SN 1954A,
SN 1993J provided an object for which high S/N ratio spectra could be
obtained over an extended period of time.  Any intrinsic small-scale
structure in the emission lines of SNe is normally lost in the
noise of the spectra.  For SN 1993J, the noise over most of our range
($3500-9000$ \AA) does not obscure the small-scale features in the
spectra until several hundred days after the explosion.  This allows
us to make an extensive study of line substructure as the supernova
evolves.  Telluric absorption could introduce spurious small-scale
structure, but we attempt to remove its effects (see Appendix).

Careful inspection of the spectra of SN 1993J reveals small-scale
structure in many emission lines, especially [\ion{O}{1}]
$\lambda\lambda$6300, 6364, [\ion{O}{1}] $\lambda$5577, \ion{O}{1}
$\lambda$7774, \ion{Mg}{1}] $\lambda$4571, and [\ion{Ca}{2}]
$\lambda\lambda$7291, 7324.  The [\ion{O}{1}] $\lambda\lambda$6300,
6364 lines in particular show a distinctive sawtooth profile similar
to those described above for SN 1987A and SN 1985F (Figure 1).  Note
that all spectra have had the systemic velocity of SN 1993J removed
($-140$ km~s$^{-1}$, determined from narrow lines in early-time
spectra; see, e.g., FMH93).  To isolate these fluctuations, we removed
the global shape of the underlying line by smoothing the profile, and
then subtracting the smoothed version.  The spectra were smoothed with
a running boxcar with a width of 60 \AA---wide enough to remove the
small-scale fluctuations from the line profile, but narrower than the
line itself, so that the subtraction would leave only the small
features.  This is the same method employed by Filippenko \& Sargent
(1989) to isolate similar substructure in the [\ion{O}{1}]
$\lambda\lambda$6300, 6364 lines of SN 1985F, although they used a
smaller boxcar width for smoothing that was better matched to the
velocity width of the fluctuations in SN 1985F.

Such a technique is similar to the process of unsharp masking in Fourier
space.  We experimented with this and other methods in the Fourier
domain to isolate high-frequency components of spectra.  The results
were not superior to the smoothing technique.  Moreover, the Fourier
techniques became less successful as the relative noise level increased
in later spectra.  The subtraction of a smoothed line from itself
does, however, introduce an artifact in the resulting residuals due to
poor removal of the edges of the global line profile.  These artifacts
have a different character than the legitimate fluctuations; they are
broader and generally show a gradual decrease from the subtracted
continuum.  Once an actual feature is reached, it clearly stands out
from the edge effect.  This is illustrated in Figure 2 with an
artificial profile from which the smoothing technique isolates the six
introduced components.  The edge effects are prominent, but distinctly
different in velocity width and character from the actual fluctuations.  To
eliminate the edge effects would require a model of the underlying
line profile.  This model would vary from line to line, and would also
change over time as different amounts of nearby lines contaminate
the feature under consideration.  For example, as illustrated in
Figure 4 of FMB94, the relative levels of [\ion{O}{1}]
$\lambda\lambda$6300, 6364 and H$\alpha$ $\lambda$6563 can change
dramatically over a time scale of months.  Subtracting a different
line profile for different spectra at different times could introduce
spurious components or temporal differences.  In the interest of
consistency, we chose to use the smoothing technique to identify
substructure in the lines.  In no case did it fail to find components
visible in the original spectra, and the only new features introduced are
the edge effects that are easily distinguishable from genuine components.

Figure 3 shows a typical example of the clumps in the [\ion{O}{1}]
$\lambda\lambda$6300, 6364 lines on day 209.  There are five
distinctive maxima marked in the figure.  S94 found six clumps, but
his clump \emph{f} (at $\sim$ 2900 km~s$^{-1}$) is difficult to
discern in this doublet; his clumps \emph{a, b, c, d,} and \emph{e}
correspond to our clumps 5, 4, 3, 2, and 1.  For the [\ion{O}{1}]
$\lambda$6300 line, these clumps are at velocities of (starting from
the blueshifted edge) $-3220$, $-2340$, $-1510$, $-310$, and 410
km~s$^{-1}$ (where 6300 \AA\ defines zero velocity).  The marked
minima (identified as 6, 7, and 8) have velocities of $-2750$,
$-1930$, and $-740$ km~s$^{-1}$, respectively.  The uncertainty in the
values for the velocities of clumps is $\sim$ 50 km~s$^{-1}$.  The
corresponding locations of these components for the [\ion{O}{1}]
$\lambda$6364 line are also indicated, but the features themselves are
difficult to distinguish.  There is also an emission feature at $\sim$
2000 km~s$^{-1}$, but it is unclear if this is from [\ion{O}{1}]
$\lambda$6300, [\ion{O}{1}] $\lambda$6364, both, or H$\alpha$.  In
fact, the large velocity width of H$\alpha$ at these times (FWHM
$\approx$ 17,000 km~s$^{-1}$, full width at zero intensity [FWZI]
$\approx$ 23,000 km~s$^{-1}$, FMB94; see also below) significantly
contaminates the spectrum to a blue wavelength of at least 6375 \AA,
possibly as far as 6310 \AA.  The presence of H$\alpha$ throughout the
region of the spectrum where one would find [\ion{O}{1}] $\lambda$6364
makes the identification of the [\ion{O}{1}] $\lambda$6364 clumps
problematic at best.

The [\ion{O}{1}] $\lambda\lambda$6300, 6364 clumps described above
first become obvious in our day 93 spectrum.  They are not apparent on
day 45.  As can be seen in Figure 4, once the features appeared, they
remain consistent until day 433.  Beyond that day, the clumps may
still be present, but the supernova had faded to the point that it was
difficult to obtain spectra with a S/N ratio high enough to
distinguish the clumps.  The clumps at $-2340$ and 410 km~s$^{-1}$
are, in fact, not obvious even in the day 433 spectrum.  There are
some slight changes in the individual clumps.  The clump at $-1510$ km
s$^{-1}$ appears to broaden slightly, from 370 km~s$^{-1}$ on day 93
(all widths are FWHM) to 470 km~s$^{-1}$ on day 167, while increasing
in amplitude in comparison with the other clumps.  The amplitude
increase continues past day 167 to the time when it became difficult
to distinguish the clumps.  The clump at $-310$ km~s$^{-1}$ grows
considerably in relative strength, by almost a factor of two from day
93 to day 433.  The clump at 410 km~s$^{-1}$ narrows from 540 km
s$^{-1}$ on day 93 to 390 km~s$^{-1}$ on day 167, after which it
remains constant.

The features seen in the [\ion{O}{1}] $\lambda\lambda$6300, 6364 lines
are also found in other oxygen emission lines.  Figure 5 compares
[\ion{O}{1}] $\lambda\lambda$6300, 6364 with [\ion{O}{1}]
$\lambda$5577 and \ion{O}{1} $\lambda$7774.  The clumps line up quite
well between [\ion{O}{1}] $\lambda$6300 and [\ion{O}{1}]
$\lambda$5577. Li et al. (1994) found a similar correspondence for a
single epoch.  The only significant differences are slight changes in
relative intensity of some of the clumps (compare the clumps at $-310$
and 410 km~s$^{-1}$ in [\ion{O}{1}] $\lambda$6300 and [\ion{O}{1}]
$\lambda$5577).  These may suggest intrinsic differences in physical
conditions in the two clumps, but are more likely the result of
contamination of the lines by other species (see above and Houck \&
Fransson [1996] for a discussion of potential line contaminants in SN
1993J).  S94's clump \emph{f} at $\sim$ 2900 km~s$^{-1}$ is apparent
in [\ion{O}{1}] $\lambda$5577 and \ion{O}{1} $\lambda$7774.  The
\ion{O}{1} $\lambda$7774 line is missing the $-2340$ km~s$^{-1}$
component completely, and the $-3220$ km~s$^{-1}$ clump is actually at
$-3110$ km~s$^{-1}$, slightly redshifted in comparison with the other
oxygen lines.  The shift of a clump is most likely the result of
contamination, but the absence of one clump from \ion{O}{1}
$\lambda$7774 that is clearly evident in [\ion{O}{1}] $\lambda$6300
and [\ion{O}{1}] $\lambda$5577 is puzzling.  The \ion{O}{1}
$\lambda$7774 line is contaminated by telluric absorption (see
Appendix), but this is not likely to result in the apparent loss of a
single component, especially not consistently over many epochs.  The
difficulty in determining actual flux values for the clumps
effectively precludes any evaluation of the differing physical
parameters for the various clumps, but the changing physical
conditions from clump to clump are most likely the cause of the slight
differences seen when comparing the various oxygen lines.  In
addition, the contamination of all the lines makes even global
evaluations highly uncertain.

There is evidence for clumps in lines of other species.  Figure 6
shows the small-scale structure of \ion{Mg}{1}] $\lambda$4571 from day
139 to day 298, while Figure 7 shows them for [\ion{Ca}{2}]
$\lambda\lambda$7291, 7324 from day 123 to day 298 (7291 \AA\ defines
the zero velocity for calcium).  (The fact that
small-scale fluctuations in magnesium and calcium lines are visible
for only a subset of the number of days that they are present in
oxygen is more indicative of the vagaries of observation than a
difference in the physical conditions producing the lines; earlier
spectra are contaminated by other lines, while later spectra are
noisier.)  Again, the features are fairly consistent over time, but,
as Figure 8 shows in a detailed comparison of [\ion{O}{1}]
$\lambda\lambda$6300, 6364, \ion{Mg}{1}] $\lambda$4571, [\ion{Ca}{2}]
$\lambda\lambda$7291, 7324, and H$\alpha$ on day 209, the pattern in
the clumps is very different between the species, although Li et
al. (1994) found that some features did correspond between H$\alpha$
and [\ion{O}{1}] $\lambda$6300.  Figure 9 shows [\ion{O}{1}]
$\lambda\lambda$6300, 6364, H$\alpha$, and \ion{Mg}{1}] $\lambda$4571
from day 433, and the differences between the lines are even more
apparent.  The H$\alpha$ line will be discussed in more detail in \S
3.  The \ion{Mg}{1}] $\lambda$4571 line does have one clump (410
km~s$^{-1}$) that lines up with oxygen, but the rest of the structure
is very different, with relatively fewer fluctuations, as in SN 1985F
(Filippenko \& Sargent 1989).  (The minimum at $-2740$ km s$^{-1}$ in
[\ion{O}{1}] $\lambda\lambda$6300, 6364 marginally lines up with a
minimum in \ion{Mg}{1}] $\lambda$4571, but it is probably
coincidental.)  This may indicate that the magnesium and oxygen
emission arise from substantially different clumps.

The [\ion{Ca}{2}] $\lambda\lambda$7291, 7324 lines have some
substructure, but they are relatively smooth.  Aside from the peaks
of the two lines and the edge effects described above, there is very
little small-scale structure that is consistent.  There is a small
peak at $\sim$ $-1700$ km~s$^{-1}$, but no other significant evidence
for persistent clumps.  In addition, these lines coincide with a
telluric water absorption feature (cf. Figure 16), so incomplete
removal of the water band profile may contaminate this region.

The scale of the clumps is not hidden by the resolution of the
observations.  The low-dispersion spectra used here have typical
resolutions of $6-7$ \AA\ (FWHM).  At 6300 \AA, that corresponds to
$\sim$ 300 km~s$^{-1}$.  Comparison with the high-resolution HIRES
spectrum illustrates that there is little, if any, substructure with
scales smaller than this (see Figure 9 of Paper I).  Virtually all the
clumps that are apparent in the HIRES spectrum appear in the
low-dispersion spectrum taken four days later.  The clumps that appear
with lower contrast in the low-dispersion spectrum, but are easily
found using our smoothing technique, do stand out more clearly in the
HIRES spectrum.  The HIRES spectrum shows that we are not missing any
of the significant details of the clumps in the low-resolution
spectra.

\subsection{Discussion of the Clumps}

The contrast between the presence of clumps in the emission lines of
oxygen (and other species) and the relative smoothness of the calcium
emission lines implies that the sources of emission for the various
species are distributed differently in the supernova ejecta.  One
possibility is that the oxygen emission arises in distinct clumps of
material while the source of calcium emission is distributed uniformly
throughout the ejecta.  While attempting to model the [\ion{O}{1}]
$\lambda\lambda$6300, 6364 lines and the \ion{Ca}{2} lines
([\ion{Ca}{2}] $\lambda\lambda$7291, 7324 and the near-infrared
[near-IR] triplet), Li \& McCray (1992, 1993) developed a similar
concept to explain the structure seen in emission lines of SN 1987A
that parallel those described above for SN 1993J.

Li \& McCray (1992) initially focused on the forbidden oxygen lines.
They developed a model based upon the relative line strengths at a
given epoch for the [\ion{O}{1}] $\lambda\lambda$6300, 6364 doublet
that predicted a value for the total oxygen mass to filling-factor
ratio.  By choosing a reasonable value for the total oxygen mass (from
theoretical models), they concluded that the oxygen emission must come
from clumps distributed uniformly throughout the ejecta with a small
filling factor.  The clumps would develop from material synthesized in
earlier evolution or during the explosion itself.  As already
mentioned, fast-moving oxygen components are seen in SNRs (Winkler \&
Kirshner 1985).  When Li \& McCray (1993) subsequently analyzed the
calcium lines, they found that the observed lines were too weak given
the probable amount of calcium produced during prior evolution and
from explosive nucleosynthesis.  Li \& McCray hypothesized that clumps
of calcium do form but do not contribute to the emission.  These
clumps of calcium intercept $\gamma$-ray radiation in proportion to
their mass fraction.  As this fraction is small, they receive only a
small amount of radioactive luminosity and thus they do not achieve
temperatures necessary for emission, although oxygen lines can be
excited at these temperatures.  The pre-existing envelope of the star
is heated generally as hydrogen and helium do not radiate efficiently,
allowing calcium that was thoroughly mixed into the atmosphere, most
likely present from the time of the progenitor's formation, to reach
the temperatures required to produce \ion{Ca}{2} emission.

This scenario of emission from smoothly distributed, pre-existing
calcium while oxygen is clumped with a small filling factor explains
the detailed line structure seen in SN 1993J.  Other attempts to model
the spectra of SN 1987A, however, concluded that the source of the
oxygen emission consisted chiefly of the envelope, and so was also the
result of pre-existing constituents (Swartz, Harkness, \& Wheeler
1989; Swartz 1991), although later analysis by the same authors
acknowledged that some oxygen from the core may contribute to the
emission and that it might be clumped (Wheeler, Swartz, \& Harkness
1993) .  Smoothly distributed oxygen is difficult to reconcile with
the clumps exhibited by the oxygen-line profiles.  The underlying
smooth line may be the result of the pre-existing oxygen, but clumps
are clearly present in SN 1993J and SN 1987A, implying that there is a
distinct difference in the site of oxygen and calcium emission.

The sawtooth appearance of the [\ion{O}{1}] $\lambda\lambda$6300, 6364
lines indicates a rough scale for the size of the clumps in the
ejecta.  Chugai (1994) used similar data for SN 1987A to develop a
technique for determining the oxygen mass.  Given the minimum cloud
size (on a velocity scale), the total velocity width of the lines, and
the relative level of fluctuation in the clumps (the scale of the
clumps compared to the overall line strength), one can then estimate
the filling factor of the clumps ($f \approx 0.1$ in the case of SN
1987A).  Several groups used the relative ratio of the [\ion{O}{1}]
$\lambda\lambda$6300, 6364 lines in SN 1987A to derive an oxygen mass
to filling-factor ratio ($M_O / f$) of 11 $M_{\sun}$ (Li \& McCray
1992) to 15 $M_{\sun}$ (Spyromilio \& Pinto 1991; Chugai 1992;
Andronova 1992).  This implies a total oxygen mass of $1.1-1.5$
$M_{\sun}$ (Chugai 1994).  As the line widths for SN 1993J are much
larger (FWHM $\approx$ 5400 km~s$^{-1}$ for SN 1993J on day 167
vs. FWHM $\approx$ 2700 km~s$^{-1}$ for SN 1987A on day 173 [Li \&
McCray 1992]), the two doublet lines are blended, as well as severely
affected by the broad H$\alpha$ line, and this effectively precludes a
new evaluation for $M_O / f$.  If one makes the admittedly dangerous
assumption that this ratio is relatively constant among core-collapse
SNe, then Chugai's method can be applied.  For SN 1993J, the sizes of
the clumps imply a typical cloud radius (in velocity space) of 150
km~s$^{-1}$ with a relative fluctuation compared to the total line
flux of $\delta\rm{F}/\rm{F} \approx 0.1$.  To estimate the effective
radius (in velocity space) for Chugai's calculation, we use Li \&
McCray's (1992) definition that emission at $|v| \lesssim v_{eff}$
constitutes 90\% of the flux in the line (Li \& McCray refer to
$v_{eff}$ as the ``expansion velocity'').  For SN 1993J, the FWHM is
$\sim$ 5400 km~s$^{-1}$, and, assuming a Gaussian profile for the
line, this implies that 90\% of the emission of the line comes from
within $v \approx \pm 3800$ km s$^{-1}$.  With $v_{eff} \approx 3800$
km~s$^{-1}$, we find a filling factor of 0.06 by Chugai's method.
Using the oxygen mass to filling-factor ratios derived for SN 1987A,
these clump parameters yield an oxygen mass of $0.7-0.9 M_{\sun}$ for
SN 1993J.

The preferred models of Nomoto et al. (1993) and Shigeyama et
al. (1990) for SN 1993J predict an oxygen mass of $\sim$ 0.4
$M_{\sun}$.  This is similar to the value of $0.47 M_{\sun}$ for
oxygen in the model of Swartz et al. (1993).  The models of Woosley et
al. (1994) and Woosley, Langer, \& Weaver (1995) yield $0.5-0.7
M_{\sun}$ for SN 1993J.  Utrobin (1994) finds a smaller value for the
oxygen mass of $\sim$ 0.2 $M_{\sun}$, but also predicts a slightly
narrower velocity profile for the oxygen lines ($v \lesssim 4600$
km~s$^{-1}$); the day 167 spectrum has a blue edge for the
[\ion{O}{1}] $\lambda$6300 line that indicates an expansion velocity
of $\sim$ 4900 km~s$^{-1}$.  For SN 1987K, another Type IIb SN,
Schlegel \& Kirshner (1989) found an oxygen mass of $\sim$ 0.3
$M_{\sun}$.  The value we derive for SN 1993J is probably a bit high,
but the blanket assumption that $M_O / f$ is constant among
core-collapse SNe definitely increases the uncertainty of the result.
A slightly smaller $M_O / f$ ratio would make our derived oxygen mass
entirely consistent with the model predictions.

\section{Late-Time Spectra}

\subsection{Late-Time Studies of Supernovae}

Few SNe have had relatively frequent spectroscopic observations for
more than three years.  Several have have been recovered many years
(even decades) after explosion.  These include SN 1957D (Long, Blair,
\& Krzeminski 1989; Cappellaro, Danziger, \& Turatto 1995), SN 1970G
(Fesen 1993), SN 1978K (Ryder et al. 1993; Chugai, Danziger, \& Della
Valle 1995; Chu et al. 1999), SN 1979C (Fesen \& Matonick 1993; Fesen
et al. 1999, hereinafter F99), SN 1980K (Fesen \& Becker 1990;
Leibundgut et al. 1991; Uomoto 1991; Fesen \& Matonick 1994; F99), SN
1985L (Fesen 1998), SN 1986E (Cappellaro et al. 1995), SN 1986J
(Leibundgut et al. 1991; Uomoto 1991), and SN 1994aj (Benetti et
al. 1998).  SN 1988Z showed evidence for circumstellar interaction
from its early-time spectra, and continues to be observed
spectroscopically with $\sim$ 10 years of coverage (Filippenko 1991a,
b; Stathakis \& Sadler 1991; Turatto et al. 1993; Aretxaga et
al. 1999).  SN 1987A represents a special case; its proximity allows
continued monitoring that would most likely not be possible for this
SN if it were in a more distant galaxy, and so we will exclude it from
this discussion\footnote{As of early 2000, though, signs of more
circumstellar interaction in SN 1987A are beginning to appear (e.g.,
Bouchet et al. 2000)}.

SNe such as SN 1988Z that show the spectroscopic characteristics of
circumstellar interaction from an early time do so through very
strong, narrow emission, especially in the Balmer lines.  They are
often referred to as a subclass of SNe II, either as ``Seyfert 1''
types or the more commonly used Type IIn (Filippenko 1989, 1991a;
Schlegel 1990).  If we ignore the SNe IIn, and consider only the SNe
II observed at very late times that appeared normal initially, then
all of the SNe recovered after many years have fairly similar spectra.
The pre-condition of normalcy at maximum brightness is not necessarily
applicable to all the SNe listed above, as some (SN 1957D, SN 1978K,
and SN 1986J) were not actually observed until much later in their
evolution.  All but SN 1978K have fairly broad emission lines (FWZI
$\approx$ 10,000 km~s$^{-1}$) of H$\alpha$, [\ion{O}{3}]
$\lambda\lambda$4959, 5007, and [\ion{O}{1}] $\lambda\lambda$6300,
6364 at late times.  Most of these lines show a largely box-like
profile, as expected from an expanding, roughly spherical, shell.

\subsection{Prior Analysis of our Nebular to Late-Time Spectra}

Spectra of SN 1993J up to day 433 are discussed by FMB94.  They
concluded that evidence for circumstellar interaction is visible as
early as day 235 when the box-like profile of H$\alpha$ became
distinct.  This shape is predicted by the circumstellar interaction
models of Chevalier \& Fransson (1994; hereinafter CF94).  Indeed,
FMB94 found an H$\alpha$ luminosity of $4.4 \times 10^{38}$
ergs~s$^{-1}$ on day 433, in approximate agreement with the models of
CF94 (and assuming the X-ray luminosity and density profile as
described in Fransson, Lundqvist, \& Chevalier [1996]).  There is a
considerable amount of evidence for circumstellar interaction in SN
1993J from observations at radio and X-ray wavelengths (see Fransson
et al. 1996, and references therein).  Houck \& Fransson (1996) and
Patat, Chugai, \& Mazzali (1995) also conclude that circumstellar
interaction is present based on late-time optical spectra.  The
spectra described here are dominated by the effects of circumstellar
interaction, and thus are discussed within that context.  These
late-time spectra are shown in their entirety in Figures 7 and 10 of
Paper I.

For our present purposes, we restrict our definition of late-time
spectra to those obtained from day 433 onward.  Nebular features begin
to appear by day 93 (cf. Figures 3, 5, and 7 of Paper I), and were
noted earlier ($\sim$ day 62) by others with better temporal coverage
during this period (Barbon et al. 1995).  From day 209, the nebular
features completely dominate the spectra, although an unusual
box-shaped H$\alpha$ line may also be apparent at that point.  This
box-shaped line is clearly visible by day 355.  From that day, the
emission characteristics of the SN begin to be dominated by evidence
for circumstellar interaction.  Nebular features in the ejecta
continue to play a significant role in the spectra, but are less
important from day 473 onward.  This section will focus on the
late-time spectra during this interaction phase, from day 473 to day
2454.  Figure 10 displays three high-quality Keck spectra of SN 1993J
from days 976, 1766, and 2454 to illustrate the discussion of the
late-time spectra.

\subsection{Extinction}

Analysis of the late-time spectra depends upon the amount of reddening
that affects them.  As absolute values for the line fluxes are
difficult to obtain, line intensity ratios must suffice, and here the
effects of reddening can be extreme.  Various values for the
extinction were found in photometric and spectroscopic studies of
SN 1993J.  Wheeler et al. (1993) used reddened blackbody models in
comparison with early-time spectra to conclude that $A_V = 0.47 \pm
0.06$ mag (all reported values of $A_V$ assume $R = A_V / E(\bv)
= 3.1$).  With similar techniques, Lewis et al. (1994) calculated $A_V
= 0.58 \pm 0.05$ mag, while Clocchiatti et al. (1995) found $A_V
\approx 0.7$ mag.  Richmond et al. (1994) summarized several
measurements of the extinction toward M81 in general and SN 1993J in
particular.  The average value is $A_V \approx 0.6$ mag, but ranges
from $A_V = 0.2$ mag to $A_V = 1.0$ mag.  Barbon et al. (1995)
estimated $A_V = 0.9$ from \ion{Na}{1}~D line equivalent widths.  The
Schlegel, Finkbeiner, \& Davis (1998) dust maps show that the Galactic
component is $A_V = 0.25$ mag, not an insignificant fraction.  For the
purposes of this discussion, we will take the extinction toward SN
1993J to be $A_V = 0.6$ mag.  All fluxes and line ratios will be
reported for the spectra as measured directly as well as corrected for
this amount of reddening using the extinction corrections of Cardelli,
Clayton, \& Mathis (1989), including the O'Donnell (1994)
modifications at blue wavelengths.

\subsection{Line Measurements}

The emission-line intensity ratios of the late-time spectra of SN
1993J are fairly difficult to measure.  The large velocity width of
the lines (FWHM $\gtrsim 15000 $ km~s$^{-1}$; see, e.g., Figures 7, 8,
10, 11, 12, and 13 of Paper I and below) and consequent overlapping
make the deblending of the lines almost impossible.  Without a clean,
isolated line to provide a well-defined profile, all measurements will
be affected by a relatively large uncertainty.  Despite this problem,
we were able to extract some information from the spectra.

The first step in the line measurement was the removal of the
underlying ``continuum.''  This is not an actual continuum from a
photosphere, but rather a mixture of blended lines that combine to
produce a relatively smooth pseudo-continuum, extending from the
ultraviolet (UV) edge of our spectra to $\sim$ 5500 \AA.  (The
pseudo-continuum was removed over the entire range of our spectra;
redward of 5500 \AA\ it is much weaker.)  The circumstellar
interaction itself can also produce a weak blue continuum (Fransson
1984).  The stronger region of the pseudo-continuum is probably iron
and iron-group element emission.  Beyond 5500 \AA, these species also
contribute, but there may be other sources responsible for the red
light.  As this is not a well-defined continuum, its removal is an
additional factor in the uncertainty of any measurements.  The
subtraction of the pseudo-continuum was accomplished with a spline fit
by hand to the spectra, but in a consistent manner for all of the
spectra to be analyzed.

Line fluxes were calculated by summing the contributions from each
pixel over a range determined by the best approximation of the edges
of the line.  In the cases of blended lines, total fluxes were
measured.  The edges of the lines provided the values for the maximum
blue and red velocities for the given line(s).  Table 1 lists the
fluxes of the lines available in our spectra relative to H$\alpha$,
with the second number indicating the line ratios when dereddened by
$A_V = 0.6$ mag.  The relative spectrophotometry of the spectra is
excellent (see Paper I for discussion of the observing and calibration
details), so the major sources of uncertainty in the fluxes are line
blending, choice of line boundaries, and the subtraction of the
continuum.  The relative flux values probably have uncertainties of
$\sim$ 10\%; individual measurements with larger uncertainties are
denoted in Table 1.  Table 2 reports the velocities of the lines
determined from the same measurements; similar caveats about the
uncertainties of the fluxes apply to the velocities, although the
values for FWHM are more secure than either the blue velocity at zero
intensity (BVZI) or the red velocity at zero intensity (RVZI).

For the day 976 spectrum (see Figure 10), we felt that the line shapes
were sufficiently simple and well-defined compared to the other epochs
to attempt a decomposition of the \ion{He}{1} $\lambda$5876 +
\ion{Na}{1}~D blend.  We modeled the line as a simple box shape with
an appropriate velocity width (15,000 km~s$^{-1}$).  Using the SPECFIT
task in the STSDAS package of IRAF, we found a good fit to the blend
with approximately equal contributions from both \ion{He}{1}
$\lambda$5876 and \ion{Na}{1}~D (Figure 11).  Later spectra were
either too noisy or did not show as clean a box shape for the profile
as did the day 976 spectrum.  Visual inspection of the later spectra,
however, indicates that the relative ratio of these two lines does not
appear to change significantly.  This can be seen in Figure 10 of Paper
I.  The blue edge of \ion{He}{1} $\lambda$5876 and the red edge of
\ion{Na}{1}~D are not as sharp in later spectra as they are on day
976.  For days 2028 to 2176, the lower S/N ratio is at least partly
responsible for degrading the line profile.  By day 2454, though, the
edges still appear, but the shoulders of the lines are more rounded.
This may indicate that the spherical structure that produces the
box-like profiles is beginning to lose its coherence.  When comparing
the relative heights of the blue and red shoulders of the \ion{He}{1}
$\lambda$5876 + \ion{Na}{1}~D blend, one must consider that the
pseudo-continuum affects the blue half more than the red.  Taking this
into account, the relative strengths of the two lines appear to remain
effectively constant.  There may be a slight increase of the strength
of \ion{He}{1} $\lambda$5876 in comparison with \ion{Na}{1}~D, but it
is not a significant change.  Therefore, the line ratio for the
\ion{He}{1} $\lambda$5876 + \ion{Na}{1}~D blend reported in Table 1
can probably be interpreted as approximately half \ion{He}{1}
$\lambda$5876 and half \ion{Na}{1}~D.

We also employed a decomposition technique that used one of the lines
in the spectra to model the others.  The individual species were
different enough that we chose to use the line of a given element to
isolate only lines from the same species, although different
ionization states were allowed.  Given the lines available, this
restricted us to the use of H$\alpha$ to model H$\beta$, and
[\ion{O}{2}] $\lambda\lambda$7319, 7330 to model [\ion{O}{1}]
$\lambda\lambda$6300, 6364 and [\ion{O}{3}] $\lambda\lambda$4959,
5007.  The line used as the model was isolated from the spectrum,
shifted in wavelength space to match the other line, and then scaled
until the subtraction of the model gave the most effective removal of
the other line.  The choice of scaling was subjective, as
contamination by other species and noisy spectra precluded a more
formal comparison.  (The [\ion{O}{1}] $\lambda\lambda$6300, 6364
doublet was removed from the blue edge of H$\alpha$ before it was
used.)  The fact that [\ion{O}{2}] $\lambda\lambda$7319, 7330 matched
the other oxygen lines so well suggests that there is very little, if
any, contamination by [\ion{Ca}{2}] $\lambda\lambda$7291, 7324 in this
wavelength region at late times.  Calcium is probably the more
significant line at earlier times, as discussed in \S 2.  The relative
line fluxes determined with this technique are reported in Table 3.
The line-subtraction technique was only effective in the spectra
from day 976 onward.  In addition, the results were more accurate for
the high S/N ratio spectra from Keck (days 976, 1766, and 2454) than
for the more noisy spectra from Lick (days 2028, 2069, 2115, and
2176).

\subsection{Discussion}

Most of our interpretation of the late-time spectra of SN 1993J falls
within a comparison with the circumstellar interaction model of CF94.
The CF94 model envisions cool, freely expanding SN ejecta colliding
with circumstellar material from a pre-explosion stellar wind.  A
forward shock propagates into the wind, while a reverse shock moves
back into the ejecta.  The SN ejecta have a fairly steep density
gradient, leading to a slow reverse shock with emission at far-UV
wavelengths (possibly in X-rays, with a different gradient).  This
produces emission from highly ionized species.  Absorption by a shell
formed at the shock boundary can yield low-ionization lines, although
these can also originate in the ejecta themselves.  CF94 treat two
different models for the structure of the wind.  One is a power law,
most applicable to a relatively compact progenitor, while the other
uses the density structure of a red supergiant (RSG) from stellar
evolution models.  They make specific predictions for their model,
including line intensity ratios and line profiles.

\subsubsection{Velocities}

The uncertainty of the edges of the lines at zero intensity, as well
as the large number of potentially overlapping lines, makes
interpretation of the measured velocities problematic.  One of the
very specific predictions of CF94's interaction model is that the line
widths decrease with time.  A pulsar-powered model would have velocity
widths that \emph{increase} slowly with time, and would have $v
\lesssim 1000$ km~s$^{-1}$ (Chevalier \& Fransson 1992).  The
velocities listed in Table 2 appear to suggest a general decrease over
time.  The H$\alpha$ BVZI is severely contaminated by [\ion{O}{1}]
$\lambda\lambda$6300, 6364 at early times, and this contamination
never completely vanishes.  In addition, H$\alpha$ may be affected on
the red side by \ion{He}{1} $\lambda$6678.  Given that \ion{He}{1}
$\lambda$5876 is approximately one-third the strength of H$\alpha$
(cf. Table 1 and \S 3.4), \ion{He}{1} $\lambda$6678 probably only has
a small effect on H$\alpha$, unless unusually high densities are
altering the typical \ion{He}{1} line intensity ratios (e.g., Almog \&
Netzer 1989).  The RVZI for [\ion{O}{3}] $\lambda\lambda$4959, 5007
is difficult to isolate, and thus is highly uncertain.  The BVZI for
H$\beta$ appears to grow fairly dramatically at late times, but this
is more likely the result of contamination by another species than any
change in H$\beta$ itself.  A comparison of H$\alpha$ and H$\beta$ in
velocity space on day 1766 and day 2454 (Figure 12), though, shows
that the blue edge of H$\beta$ does not look like H$\alpha$ at late
times and is probably some other species.  The CF94 model does predict
a strong \ion{Mg}{1}] $\lambda$4571 line, but this feature near
H$\beta$ is at the wrong wavelength to be magnesium.

The BVZI of \ion{He}{1} $\lambda$5876 and the RVZI of \ion{Na}{1}~D
are more likely to reflect the actual limits of velocity space
occupied by their respective species.  The most accurate velocity
information comes from the FWHM of H$\alpha$.  This is much less
affected by errors in the subtraction of the pseudo-continuum, and
there is less impact by weaker lines.  There is still some
contamination by [\ion{O}{1}] $\lambda\lambda$6300, 6364, but the
line-subtraction technique described above indicates that it widened
H$\alpha$ minimally.  Perhaps the best evidence that the FWHM of
H$\alpha$ is fairly accurate is how remarkably constant the values
are.  Probably from day 881, and definitely from day 1766, the FWHM is
$\sim$ 15,000 km~s$^{-1}$, within our measurement uncertainty.  The
decrease found by day 2454 appears to be real.  If this is the case,
then the line widths are decreasing in accordance with the CF94
model.  Marcaide et al. (1997) also found an apparent deceleration in
the ejecta using very long baseline interferometry.

The CF94 model has other predictions relevant to the observed
velocities.  One is that highly ionized lines should have different
velocity widths than low-ionization lines.  Unfortunately, we cannot
test this predication, as the difference is only $\sim$ 830
km~s$^{-1}$ in the power-law model for the wind structure.  Another is
the scale of the line-profile widths.  In the power-law model, the
half width at zero intensity (HWZI) is $\sim$ 5000 km~s$^{-1}$, while
for the RSG model, the HWZI is $\sim$ 4200 km~s$^{-1}$.  For the lines
of SN 1993J, the HWZI is $\sim$ 7500 km~s$^{-1}$.  This is not
surprising as the structure of the progenitor of SN 1993J would not
have been that of a typical RSG.  It may indicate that the particular
power law used is not applicable to SN 1993J either.  The four SNe
discussed by F99 (1986E, 1980K, 1979C, and 1970G) have expansion
velocity widths (HWZI) on the order of 5500 km~s$^{-1}$, implying that
they have higher-mass envelopes than SN 1993J.  This is
understandable, as most models for SN 1993J postulate that it had a
low-mass ($0.2-0.5 M_{\sun}$) hydrogen envelope, so a smaller mass
implies a larger velocity for a given amount of input energy (see
Paper I and references therein).  In all five cases, though, the
velocities are larger than predicted by any of CF94's models.  In
addition, the mass-loss rate of SN 1993J as determined from radio
observations (Van Dyk et al. 1994) is $(2-15) \times 10^{-5}
M_{\sun}$~yr$^{-1}$, and this is comparable to the values for the
other four SNe [($2-19) \times 10^{-5} M_{\sun}$~yr$^{-1}$; F99, and
references therein].

\subsubsection{Emission Lines}

Table 4 lists the predicted line fluxes from the CF94 model, both for
the power-law wind structure and the RSG model.  The measured fluxes
from our data range from 1.8 to 6.7 years, thus providing a comparison
with several epochs of the model.  Table 4 only contains the lines
with which we can make any comparisons.  The limitations of the
spectra in wavelength coverage and quality at red wavelengths restrict
the set of usable lines.  The blending of the lines also complicates
some direct comparisons.

The spectra from days 653, 670, 881, and 976 are closest to the CF94
year-2 predictions.  For these days, assuming \ion{Na}{1}~D is half
the blend of \ion{He}{1} $\lambda$5876 + \ion{Na}{1}~D as described
above, the sodium line is nearly half the predicted strength.  The
[\ion{O}{1}] $\lambda\lambda$6300, 6364 doublet is only a quarter of
the predicted value and [\ion{O}{2}] $\lambda\lambda$7319, 7330 is
three times stronger than in the model, but the [\ion{O}{3}]
$\lambda\lambda$4959, 5007 strength may be comparable.  If the line
subtraction technique is correct, and H$\beta$ is $0.2-0.3$ of the
H$\beta$ + [\ion{O}{3}] blend, then the observed oxygen-line flux is
similar to the predicted value.  Unfortunately, the [\ion{Ne}{3}]
$\lambda$3968 line was not available in all our spectra.  Its value on
day 653 is three times larger than predicted, but this region has the
most uncertainty due to potential line contamination and the
decreasing quality of the spectra toward blue wavelengths.

The later spectra are closer in time to the year-5 model of CF94, but
we will also consider the year-10 models.  The \ion{Na}{1}~D strength
is still weaker than predicted, but closer to calculated values, and
would actually fit well with the prediction for year 2.  The relative
strength of the sodium line does increase, as is the general trend in
the models.  The RSG model actually predicts a quite low value for
\ion{Na}{1}~D, in contrast to the power-law model.  While continuing
to gain in relative strength, [\ion{O}{1}] $\lambda\lambda$6300, 6364
is still far below the predicted values for the power-law model.  In
this case, the oxygen doublet is in agreement with the RSG value at
year 10.  The biggest difference between the observations and the
models may be the strength of [\ion{O}{2}] $\lambda\lambda$7319, 7330.
In the models, this feature barely contributes, while the observations
show it to be a significant line, with $70-80\%$ of the strength of
H$\alpha$.  Unfortunately, the CF94 RSG model does not report the
predicted strength of [\ion{O}{2}] $\lambda\lambda$7319, 7330.  The
RSG model predicts an even higher strength for [\ion{O}{3}]
$\lambda\lambda$4959, 5007 relative to H$\alpha$ than the power-law
model, and this is not what is observed, so we will concentrate on the
predictions of the power-law model.

As [\ion{O}{2}] $\lambda\lambda$7319, 7330 is expected to originate in
the ionized ejecta along with [\ion{O}{3}] $\lambda\lambda$4959, 5007,
it is unclear why the singly ionized oxygen lines should be so strong,
when the doubly ionized lines are slightly weaker than the models
indicate unless the ionization structure is fairly different from the
models assumptions.  The density may explain the ratios, as will be
discussed below.  Chevalier \& Fransson (1989) show that [\ion{O}{2}]
$\lambda\lambda$7319, 7330 is stronger in models of typical late-time
spectra of core-collapse SNe without circumstellar interaction.
Unfortunately, [\ion{Ca}{2}] $\lambda\lambda$7291, 7324 is predicted
to be an order of magnitude stronger than the [\ion{O}{2}] lines, so
the circumstellar model is more appropriate, although the depletion of
calcium onto dust grains formed as the ejecta cool may affect the
[\ion{Ca}{2}] $\lambda\lambda$7291, 7324 line strength (e.g., Kingdon,
Ferland, \& Feibelman 1995).

The [\ion{Ne}{3}] $\lambda$3968 line is again uncertain in these very
late-time spectra, but it appears to be significantly stronger than
the models predict.  This makes the issue of ionization structure even
more complicated, as this high-ionization line also originates in the
ejecta.  If the relative strengths of [\ion{O}{2}] and [\ion{O}{3}] are
the result of a lower ionization of the ejecta, then the strength of
[\ion{Ne}{3}] is difficult to explain.  This may imply that there are
also abundance issues that affect the relative line strengths.  As SN
1993J was assumed to have lost most of its hydrogen envelope (see
Paper I and references therein), perhaps the relative amount
of hydrogen has been depleted in the ejecta compared to the abundances
used in CF94's models, which were computed for normal SNe II.  This
could explain the apparent excess of emission from the lines that
originate in the ejecta ([\ion{O}{2}], [\ion{Ne}{3}]), while the lines
that originate in the shell (\ion{Na}{1}~D, [\ion{O}{1}]) seem too
weak, as all the line ratios are relative to H$\alpha$.  The strength
of [\ion{O}{3}] actually fits with the predictions, and thus does not
necessarily support this scenario.

Some of the lines predicted by CF94 are notable by their absence from
our spectra.  These include [\ion{C}{1}] $\lambda$8729, [\ion{N}{2}]
$\lambda\lambda$6548, 6583, [\ion{O}{2}] $\lambda\lambda$3726, 3729,
\ion{Mg}{1}] $\lambda$4571, and [\ion{S}{3}] $\lambda\lambda$9069,
9532.  The carbon line would not be particularly strong, but the
near-IR sulfur lines would be comparable in strength to H$\alpha$.
Neither of these features is apparent, although hints of the sulfur
line may exist in the day 2176 spectrum (Figure 10 of Paper I), but
clearly at a lower strength.  The [\ion{O}{2}] $\lambda\lambda$3726,
3729 line could be contributing to the [\ion{Ne}{3}] $\lambda$3968
line, but the distinctive double-horned oxygen line profiles described
below do not appear and the wavelengths do not match well.  As
discussed above, the \ion{Mg}{1}] $\lambda$4571 line might be
appearing at the blue edge of H$\beta$, but the velocities are not
correct.  In addition, the magnesium line should be two-thirds of the
strength of H$\alpha$, and that is obviously not the case.  The
nitrogen lines could be hidden within the H$\alpha$ structure, but
there should then be a shoulder on the blue edge of H$\alpha$ caused
by the $\lambda$6548 component.  The predicted total [\ion{N}{2}]
$\lambda\lambda$6548, 6583 flux is about one-third that of H$\alpha$,
and if the line ratios are not affected by density considerations,
then the $\lambda$6548 line may be too weak to show up on the
H$\alpha$ profile.  As we show below, there is contamination on the
blue side, but from [\ion{O}{1}] $\lambda\lambda$6300, 6364.

Most of these apparently missing lines have one attribute in
common---a low critical density.  For [\ion{O}{2}]
$\lambda\lambda$3726, 3729, [\ion{N}{2}] $\lambda\lambda$6548, 6583,
and [\ion{S}{3}] $\lambda\lambda$9069, 9532, the critical densities
are all less than $10^6$ cm$^{-3}$ for $T \approx 10,000$ K. (These,
and subsequent values for the critical density and other line
diagnostics, were calculated with the IRAF/STSDAS NEBULAR suite of
tasks.)  The [\ion{C}{1}] $\lambda$8729 line has a high critical
density ($> 10^7$ cm$^{-3}$), so its absence, along with the lack of
\ion{Mg}{1}] $\lambda$4571, may be explained by an origin in the
shell, whose lines appear to be weaker than those from the ejecta.
The critical density for [\ion{O}{2}] $\lambda\lambda$7319, 7330 at $T
\approx 10,000$ K is $5.7 \times 10^6$ cm$^{-3}$, while it is only
$6.2 \times 10^5$ cm$^{-3}$ for [\ion{O}{3}] $\lambda\lambda$4959,
5007.  This may explain the relative weakness of [\ion{O}{3}], if the
absence of the other lines does imply a density $\gtrsim 10^6$
cm$^{-3}$.

The [\ion{O}{3}] lines may be able to further constrain the density.
The [\ion{O}{3}] $\lambda$4363 and $\lambda\lambda$4959, 5007 lines
are contaminated by H$\gamma$ and H$\beta$, respectively.  This
contamination is potentially significant from days 553 to 1766, but
probably less than 20\% thereafter (cf. Tables 1 and 3).  The relative
contamination of H$\beta$ would be greater, and this would tend to
drive the [\ion{O}{3}] ($\lambda\lambda$4959, 5007/$\lambda$4363)
ratio lower as it dilutes [\ion{O}{3}] $\lambda\lambda$4959, 5007 if
removed properly.  Residual H$\beta$ contamination could drive the
[\ion{O}{3}] ($\lambda\lambda$4959, 5007/$\lambda$4363) ratio higher.
Without considering the effects of the hydrogen lines, though, the
[\ion{O}{3}] ($\lambda\lambda$4959, 5007/$\lambda$4363) ratio is
already $\sim$ 1 at day 670, implying that, to within the
uncertainties of our measurements, we can probably ignore the hydrogen
without affecting our interpretation significantly.  To get
[\ion{O}{3}] ($\lambda\lambda$4959, 5007/$\lambda$4363) as low as 1,
the temperature must be at least 10,000 K, even at the large densities
considered here.  The electron density under these conditions would be
$n_e \gtrsim 10^8$ cm$^{-3}$.  By day 2454, the ratio is $\sim$ 3,
implying a density of $n_e \approx 10^7$ cm$^{-3}$.  For higher
temperatures, the required density drops.  If $T \approx$ 30,000 K,
for example, $n_e \approx 10^7$ cm$^{-3}$ on day 670 and $n_e \approx
2 \times 10^6$ cm$^{-3}$ by day 2454.  No matter what the temperature
is, the increase of the [\ion{O}{3}] $\lambda\lambda$4959, 5007 to
$\lambda$4363 ratio with time implies a decreasing density, as one
would expect in the expanding ejecta.  Given the discussion of
critical densities above, a density $\gtrsim 10^6$ cm$^{-3}$ with a
temperature of $T \approx 10,000-30,000$ K appears to be consistent
with the line emission.

The relative ratios observed are, for the most part, consistent with
values determined by F99 from other late-time spectra of core-collapse
SNe.  Both SN 1980K and SN 1979C have very similar line ratios when
compared with SN 1993J.  This includes the large amount of emission in
[\ion{O}{2}] $\lambda\lambda$7319, 7330.  SN 1986E has a much weaker
[\ion{O}{2}] feature (see also Cappellaro et al. 1995; they identify
the line at $\sim$ 7300 \AA\ as the [\ion{Ca}{2}]
$\lambda\lambda$7291, 7324 doublet).  SN 1979C is the only one of
these four to have a measured [\ion{Ne}{3}] $\lambda$3968 line, and it
is not as strong as in SN 1993J.  The lines predicted by CF94 that are
missing in SN 1993J are also missing in these SNe, suggesting that
they, too, have a density higher than the critical density of these
lines.

A few of the late-time spectra were observed on photometric nights.
For these, we calculated the luminosity of the H$\alpha$ line using a
distance to M81 of $d = 3.6$ Mpc (Freedman et al. 1994).  On days 553,
670, and 881, the H$\alpha$ luminosity was (in units of $10^{38}$
ergs~s$^{-1}$) 2.5, 3.0, and 4.8, respectively.  By day 1766, it was
0.86, and 0.37 on day 2454.  (To correct for an extinction of $A_V =
0.6$ mag, the luminosities would be multiplied by 1.5.)  Both the day
553 and day 670 spectra were observed through relatively narrow slits
on nights with variable seeing, implying a larger uncertainty for the
luminosities calculated for those observations.  The day 881 spectrum
was taken with a 4$\arcsec$ slit width; thus, the value of 4.8 $\times
10^{38}$ ergs~s$^{-1}$ is probably a better indication of the
H$\alpha$ luminosity during these days.  Note that FMB94 found the
luminosity of the H$\alpha$ line to be $4.4 \times 10^{38}$
ergs~s$^{-1}$ on day 433---a comparable value within our probable
uncertainties.  The day 1766 and day 2454 spectra were also taken
through narrow slits, with the seeing comparable to the slit width.
Therefore, the luminosities at these later times could be larger (or
smaller) than measured, but most likely by not more than a factor of
two, implying a real decrease in H$\alpha$ luminosity by such late
times.  CF94 predicted an H$\alpha$ luminosity of 0.45, 0.19, and 0.09
$\times 10^{38}$ ergs~s$^{-1}$ at 2, 5, and 10 years, respectively,
for the power-law model.  The RSG model gave an H$\alpha$ luminosity
of 0.48 $\times 10^{38}$ ergs~s$^{-1}$ at 10 years.  Patat et
al. (1995) found an H$\alpha$ luminosity of $1.3 \times 10^{38}$
ergs~s$^{-1}$ (corrected for extinction of $A_V = 0.1$ mag) on day 368
(their day 367) and they interpreted this level of emission as a
requirement for circumstellar interaction; radioactive decay alone was
not enough.  Our values are consistent with the Patat et al. (1995)
model that requires circumstellar interaction to power the emission.
The luminosities we observe at late times for SN 1993J agree with the
values reported for other SNe observed at similar times (see F99 for a
table of values).

\subsubsection{Overall Comparison with the Circumstellar Interaction
Model}

The CF94 model makes many predictions for the spectrum resulting from
circumstellar interaction.  The comparison of these predictions with
the late-time spectra of SN 1993J and the prior study of F99 indicates
that the model is somewhat successful, but needs refinement.  The
velocity widths in SN 1993J and the four SNe of F99 are larger than
the model predicts, especially larger than those of the RSG model, but
they do follow the expected trend of a general decrease with time.
The line intensity ratios of the RSG model seem fairly inconsistent
with the observed values.  The power-law model is slightly better, as
the trends of the relative ratios over time are consistent.  One
exception to this is the prediction of a strengthening of H$\beta$
relative to H$\alpha$ as the interaction continues; we do not see this
in SN 1993J (cf. Tables 3 and 4), although the density of the emitting
material may affect this result.  The trend of a decreasing luminosity
for the H$\alpha$ line at late times is also consistent between the
observations and the models, although the absolute values differ.

The power-law model predictions may not be so different from the
observed results if density effects are considered.  Many of the
emission lines predicted by the model that have low critical densities
are absent from the spectra of SN 1993J and the four SNe discussed by
F99.  SN 1993J may have a relatively lower hydrogen abundance, as
shown by its transformation from Type II to Type IIb.  This will
increase the relative intensity of other lines.  The fact that similar
values are seen in the four SNe of F99, though, implies that the
low-mass envelope of SN 1993J is not altering the circumstellar
interaction significantly.  A larger density for the emitting region
associated with the circumstellar interaction may be the main
explanation for the deviation of the line intensity ratios from the
predictions of the CF94 models.  The relative weakness of lines
predicted to arise in the ejecta, as opposed to the shell, argues that
a depleted level of hydrogen may have some impact on the
line-intensity ratios.

\subsubsection{H$\alpha$ Line Structure}

One other aspect of the late-time H$\alpha$ line is that it shows a
persistent, narrow (unresolved, FWHM $\lesssim$ 250 km~s$^{-1}$ on day
1766) feature at zero velocity (see Figure 9 for a detailed view on
day 433).  Careful inspection of Figure 10 shows that the line is
present in all three of the high S/N ratio, late-time spectra.  It is
not clear if this is emission from flash-ionized circumstellar
material around SN 1993J (as noted in early spectra, see Paper I and
references therein), or merely an \ion{H}{2} region along the line of
sight.  Finn et al. (1995) noted the presence of this feature and
found it to be slightly variable in strength.  They also show (their
Figure 7) an image of the environment of SN 1993J taken through a
narrow-band filter centered on H$\alpha$.  There is faint \ion{H}{2}
region emission evident around the SN.  Aldering, Humphreys, \&
Richmond (1994), while studying earlier images of M81 to analyze the
progenitor of SN 1993J, found that the colors indicated more than one
star, and perhaps a faint OB association, that could be the source of
the narrow hydrogen emission.  Finn et al. (1995) concluded that the
feature was probably not directly connected with SN 1993J.  If the
emission is not associated with SN 1993J itself, then it cannot
originate very far from the line of sight to SN 1993J.  Data from the
Keck telescope on day 1766 were taken through a 1$\arcsec$ slit with
$\sim$ 1$\arcsec$ seeing (the extraction width was 4.$\arcsec$2).  The
narrow feature is still present, and thus is likely directly along the
line of sight.  The H$\beta$ line shows no evidence for this feature,
but the S/N ratio of the spectra in this region would make
identification difficult.

Another possible source for the narrow H$\alpha$ line is material
ablated from a companion star.  As discussed in Paper I, many models
for SN 1993J postulate a companion to facilitate the loss of the
hydrogen envelope from the progenitor.  Chugai (1986), Livne, Tuchman,
\& Wheeler (1992), and Marietta, Burrows, \& Fryxell (2000) have
considered the effects of the interaction of a SN explosion with a
companion.  Their calculations indicated that material from a
companion to an exploding white dwarf could be swept up in the ejecta
with emission becoming visible several hundred days after the
explosion.  The predicted widths, though, range from a few hundred to
over a thousand kilometers per second.  Given the low velocity width
of the observed late-time narrow H$\alpha$ line in SN 1993J, it is
unlikely that ablated material from the companion star is the source.

As can be seen in Figure 10, there are small-scale features covering
the flat top of the H$\alpha$ profile similar to the clumps described
in \S 2 for the oxygen lines.  In this case, though, the structures
change over time.  As noted by FMB94, they probably represent
Rayleigh-Taylor instabilities in the hydrogen-emitting region behind
the shock (Chevalier, Blondin, \& Emmering 1992).  The instabilities
are enhanced if the shock is radiative (Chevalier \& Blondin 1995), as
Fransson et al. (1996) suggest for SN 1993J.

\subsubsection{Late-Time Oxygen Lines}

The relative strengths of the various oxygen lines are shown by the
ratios determined using the line-subtraction technique described in \S
3.4 and presented in Table 3.  The [\ion{O}{1}] $\lambda\lambda$6300,
6364 to [\ion{O}{2}] $\lambda\lambda$7319, 7330 ratio is approximately
constant.  The values for the ratio of [\ion{O}{3}]
$\lambda\lambda$4959, 5007 to [\ion{O}{2}] $\lambda\lambda$7319, 7330
are difficult to interpret given their apparently large and random
range.  The values from the Keck spectra (days 976, 1766, and 2454)
are the most reliable, and, considering only these values, the
relative strength of [\ion{O}{3}] $\lambda\lambda$4959, 5007 appears
to weaken slightly compared with [\ion{O}{2}] $\lambda\lambda$7319,
7330 over the course of the observations.  This may indicate a
reduction in the strength of the interaction and thus a decrease in
the ionization level within the ejecta.  Given the uncertainty in
these values, this is highly speculative.

A striking change occurs in the spectra of SN 1993J between days 976
and 1766.  The oxygen lines ([\ion{O}{1}] $\lambda\lambda$6300, 6364,
[\ion{O}{2}] $\lambda\lambda$7319, 7330, and [\ion{O}{3}]
$\lambda$4363 and $\lambda\lambda$4959, 5007) all develop narrower
profiles on top of the underlying box-like shape (cf. Figure 10).
These features are not contamination by other lines, as they appear
with the same relative velocities for all of the oxygen lines
accessible in our spectra (Figure 13).  The peaks of the narrower
profiles are separated by $\sim$ 7100 km~s$^{-1}$, so these are not
the two components of the oxygen doublets (the maximum velocity
separation for the oxygen doublets is $\sim$ 3000 km~s$^{-1}$ for
[\ion{O}{1}] $\lambda\lambda$6300, 6364).  We believe that these
profiles represent the blue and red peaks of a double-horned profile
due to the ejecta colliding with a disk-like, or at least a somewhat
flattened, region.  (For a derivation of the two-horned profile from
an expanding disk-like structure, see, e.g., Leonard et al. 2000;
Gerardy et al. 2000.)  The disk-like structure may be due to the
interaction of circumstellar material with a binary companion, thought
to be present according to models for mass loss from the progenitor of
SN 1993J (see Paper I and references therein).  The red horn is weaker
than the blue horn due to differential extinction; we see the former
through a much longer path in the ejecta.  The large gap in our
coverage between days 976 and 1766 means that the onset of the disk
features was not recorded.  There are vague hints of the incipient
two-horned structure on day 976 (Figure 14), but it is certainly not
as significant as in the later spectra.  This type of structure is not
evident in any of the four SNe whose late-time spectra are discussed
by F99.

Starting with the day 1766 spectrum, a notch develops in the middle of
the H$\alpha$ line.  This loss of flux at zero velocity may indicate a
slight flattening of the spherical distribution that had been the
source of the hydrogen emission.  Blondin, Lundqvist, \& Chevalier
(1996) present a model for an axisymmetric circumstellar interaction
in which the SN is interacting with a dense equatorial structure.
Such a scenario would result in little emission along the axis of
symmetry because of a lower abundance of circumstellar material, while
the equatorial structure would emit as in a typical interaction.  This
could create a disk-like or flattened source of circumstellar emission
as is seen here.  As the ejecta expand, they can interact with the
disk-like or flattened circumstellar material producing the two-horned
profiles.  The emission-line profile from an expanding disk is double
peaked and resembles that of a bipolar ejection.  In fact, some new
models for the core-collapse mechanism may require a bipolar outflow
(Khokhlov et al. 1999), although they are fairly speculative.  In
addition, the optical spectropolarimetric studies of SN 1993J
indicated asymmetries at early times (Trammell, Hines, \& Wheeler
1993; Tran et al. 1997).  Polarimetry of other SNe (e.g., Leonard et
al. 2000; Wang et al. 2000) also suggests that asymmetries might be
common, although the late appearance of a non-spherical emission
source may indicate that the circumstellar matter is the cause of the
asymmetry.

The appearance of this disk-like structure may complicate the
interpretation of the line-intensity ratios discussed above.  There is
still the box-like profile from a roughly spherical source, in
addition to the flattened region.  The discussion of line ratios
assumed that the conditions were the same for the entire emitting
region.  It may be that the shell and the disk have distinctly
different physical conditions, but such differences would be very
difficult to separate in our spectra.  The wings of the disk-like
profile are obvious, but the precise decomposition of the line into
disk and shell components would be model dependent.  This is made even
more complicated by the fact that the box-like profile shows no
significant asymmetry, while the two-horned profile is clearly
stronger in the blue component.  This implies that the flattened
region suffers from differential extinction, while the roughly
spherical shell does not.

\section{Conclusions}

The detailed substructure of the ejecta revealed by the spectra from
early to nebular times indicates that SN 1993J is a clumpy supernova.
(For a discussion of substructure that is the result of telluric
absorption, see the Appendix.)  Both oxygen and magnesium lines show
clumps.  The oxygen clumps appear in both permitted and forbidden
lines, with some slight variations.  The calcium lines do not seem to
have clumps.  This is consistent with a scenario in which the observed
oxygen emission originates in clumps of the newly synthesized material
(either through prior evolution or explosive nucleosynthesis), while
the calcium emission arises mostly from pre-existing material
distributed uniformly throughout the envelope, following the models of
Li \& McCray (1992, 1993).  Using the technique of Chugai (1994), we
find a total oxygen mass based on the observed clumps of $M_O \approx
0.7-0.9 M_{\sun}$.  This value is very uncertain and it is larger than
the values predicted by models of SN 1993J.

We also present late-time spectra of SN 1993J.  The lines clearly
exhibit circumstellar interaction through their box-like emission
profiles.  They indicate a gradual decrease in velocity width from day
433 to day 2454, with the FWHM of H$\alpha$ remaining relatively
constant from day 976 to day 2176 at $\sim$ 15,000 km~s$^{-1}$,
decreasing slightly by day 2454.  The line flux for H$\alpha$ is
approximately constant from day 433 to day 881.  It decreases by about
a factor of five by days 1766 and 2454.  Other strong lines in the
spectrum include [\ion{O}{2}] $\lambda\lambda$7319, 7330, [\ion{O}{3}]
$\lambda$4363, [\ion{O}{3}] $\lambda\lambda$4959, 5007, [\ion{Ne}{3}]
$\lambda$3968, \ion{He}{1} $\lambda$5876 + \ion{Na}{1}~D (probably
each of equal strength), H$\beta$, and [\ion{O}{1}]
$\lambda\lambda$6300, 6364.  These lines, and their relative ratios,
agree fairly well with other studies of interacting SNe at late times
(e.g., F99).  While the circumstellar interaction model of CF94
predicts some lines that do not appear, and that [\ion{O}{2}]
$\lambda\lambda$7319, 7330 should be weak, the model is somewhat
consistent with the general line-intensity ratios, and the predicted
evolution of the line ratios agrees with the observations.  Our
spectra, though, indicate that the density is probably higher than
that used by CF94.  This may be because SN 1993J is unusual, but it is
inconsistent with both the power-law model that assumes a compact
progenitor and the RSG model that has a more extended progenitor.
Considering SN 1993J and the four SNe of F99, the power-law model
appears to be a better match in its predictions, although a higher
density may be required to reproduce accurately the observed
line-intensity ratios.

The late-time oxygen lines imply a relatively high density ($n_e
\approx 10^7-10^8$ cm$^{-3}$) on day 670, decreasing to $n_e \approx
10^6-10^7$ cm$^{-3}$ by day 2454, assuming a temperature of 10,000 K.
The low ratio [\ion{O}{3}] ($\lambda\lambda$4959, 5007/$\lambda$4363)
$\approx$ 1 at early times excludes temperatures much below 10,000 K.
The increase of this ratio to $\sim$ 3 on day 2454 shows the general
expansion of the oxygen-emitting region.  The appearance of the
two-horned profile on the oxygen lines by day 1766 exhibits the
development of emission from a somewhat flattened, disk-like structure
in the interaction region.  This may indicate a pre-existing
equatorial belt of circumstellar material around SN 1993J, possibly
similar to that observed in SN 1987A (e.g., Panagia et al. 1996, and
references therein), although the emission in SN 1993J is likely the
result of the ejecta colliding with the equatorial belt to produce the
emission.


\acknowledgments This research was supported by NSF grants AST-9115174
and AST-9417213 to A.V.F as well as by NASA through grants GO-7434,
GO-7821 and GO-8243 from the Space Telescope Science Institute, which
is operated by AURA, Inc., under NASA contract NAS 5-26555.  We are
grateful to the staffs of the Lick and Keck Observatories for their
assistance with the observations.  The W. M. Keck Observatory is
operated as a scientific partnership among the California Institute of
Technology, the University of California, and NASA. The Observatory
was made possible by the generous financial support of the W. M. Keck
Foundation.

\appendix

\section{Telluric Absorption}

One aspect of spectroscopy that is often overlooked in the study of
SNe is the impact of weak telluric absorption, which can sometimes
mimic intrinsic absorption features in SN spectra.  Virtually all
observers are familiar with the prominent A-band ($\sim$ 7600 \AA) and
B-band ($\sim$ 6850 \AA) absorptions produced by molecular oxygen, and
some make an attempt to remove them.  There are also several strong
water absorption bands in the near-IR region.  Weaker water bands in
the range $5000-6700$ \AA, however, are much less well known.  Typical
SN spectra are often so noisy that the weaker absorptions are not
obvious.  This is especially true at blue wavelengths where detectors
are usually less efficient; for example, there are ozone absorption
features in the region $3200-3450$ \AA\ (the Huggins bands; e.g.,
Schachter, Filippenko, \& Kahn 1989; Schachter 1991, and references
therein).  An extremely weak absorption, the Chappuis ozone band,
extends from 5000 \AA\ to 7000 \AA, but it is unlikely to affect SN
spectra significantly (see, e.g., Kondratyev 1969).  Even in spectra
with a moderate S/N ratio, some of the weaker bands can appear at high
airmass or high humidity.  The fortuitous appearance of a few
relatively bright SNe in the past decade, however, shows that these
features will be present even at low airmass in the spectra of all
objects.

Figure 15 displays spectra of three bright SNe (SN 1993J, SN 1994D, and SN
1999em) over the range $6360-6700$ \AA.  Each appears to contain an
absorption line at $\sim$ 6515 \AA, but the spectra are shown with
\emph{observed} wavelengths, and these three SNe have differing
redshifts ($-140$ km~s$^{-1}$, 850 km s$^{-1}$, and 720 km~s$^{-1}$,
respectively), clearly implying a local source for the absorption.
Figure 16 is a plot of the telluric absorption spectrum as determined
from solar observations obtained with the Fourier-transform
spectrometer at the McMath-Pierce telescope at Kitt Peak National
Observatory\footnote{NSO/Kitt Peak FTS data used here were produced by
NSF/NOAO.} (Wallace, Hinkle, \& Livingston 1993, 1998).  This telluric
absorption spectrum has had the amount of water absorption increased
by a factor of 1.7 to match better the conditions at Lick Observatory
(G. Marcy, 1999, personal communication), and it has been smoothed to
correspond to a spectral dispersion of $\sim$ 2 \AA/pixel.  The inset of
Figure 16 shows the telluric absorption in the region $6250-6670$ \AA.
Note that at this resolution a strong, narrow absorption appears at
6515 \AA\ within a broader water band.

The presence of the absorption at 6515 \AA\ and other, weaker features
near H$\alpha$ indicates that great caution must be taken when
interpreting small-scale fluctuations in spectra near this wavelength
range.  This telluric line was noted and discussed as a possibly
intrinsic feature in SN 1993J by Wheeler \& Filippenko (1996) and Finn
et al. (1995), although neither group attributed any significance to
it.  Since even experienced spectroscopists can be fooled by this
line, it is important to be reminded of the potential effects of weak
telluric absorption.  This is especially true in efforts to search for
the presence of H$\alpha$ in spectra of SNe Ia.  A bright SN Ia would
be an obvious choice for such an analysis, but a brighter object is
more likely to show the effects of telluric absorption and lend itself
to misinterpretation.  The presence or absence of hydrogen in emission
or absorption in the spectra of SNe Ia could have tremendous
significance in the evaluation of their proposed progenitor systems
(see, e.g., Branch et al. 1995).  One must therefore be very careful
with SN data to avoid the pitfalls of telluric absorption.  With the
development of ever larger telescopes, this problem will extend to
even fainter SNe.  In addition, the (relatively) strong O$_2$
absorption near $6260-6340$ \AA\ can appear to be an intrinsic SN
feature in spectra with a low S/N ratio.

It is possible to remove most of the effects of telluric absorption,
especially at high spectral resolution.  The division by the spectrum
of a comparison star that is intrinsically featureless in the
wavelength regions affected by absorption can correct object spectra
fairly well (Wade \& Horne 1988; Bessell 1999).  It can sometimes be
difficult to find a star that is sufficiently featureless at the
wavelengths of interest.  This technique requires a well-exposed star
to minimize the addition of any noise to the object spectrum.  It is
preferable to have a comparison star that is observed at an airmass
similar to the object, but it is possible to scale the comparison star
to match.  We use our own procedures to remove telluric absorption
following the technique of Wade \& Horne (1988)\footnote{We note that
the IRAF task TELLURIC divides the object by the comparison star
properly, but does not scale the spectrum of the comparison star
according to the method of Wade \& Horne (1988).  With both TELLURIC
and Wade \& Horne, the comparison star is exponentiated by a function
of the ratio of the airmasses of the object and the comparison star.
In TELLURIC, however, this function is the ratio of airmasses
multiplied by a scale factor, while with Wade \& Horne it is the ratio
of the airmasses raised to a power.  Wade \& Horne choose a power as
the strongest telluric lines are saturated and the equivalent widths
will grow with approximately the square root of the airmass.  The
TELLURIC scaling of the ratio of the airmasses effectively assumes an
optically thin atmospheric absorption.  When the airmasses of object
and comparison star are similar, this different treatment has little
impact.  With the Wade \& Horne method, variation of the power
depending on the source and level of saturation of the absorption may
result in better removal.  We are conducting tests in this area.}.

In the normal course of data reduction, we remove the stronger
telluric absorptions, but we only include the regions of weaker bands
when observing at high airmass (or sometimes when the humidity is
high).  This is to avoid adding undue noise in these regions when the
weak bands are already hidden in the noise of the spectrum.  High S/N
ratio objects, though, will show the weak bands.  When we make an
extra effort to remove them, as shown in Figure 15 for the 6515 \AA\
line in SN 1999em, we achieve some success.  There are still residuals
at 6515 \AA; it appears unlikely that any telluric absorption removal
scheme can be 100\% effective, but it can be done to a fairly good
level, as long as interpretation of the spectra includes consideration
of its effects.  Observers must make sure that only real lines are
identified as such.


\clearpage



\figcaption{Spectrum of the [\protect\ion{O}{1}] $\lambda\lambda$6300,
6364 blend on day 209.  Note the sawtooth structure superposed on the
global line profile.\label{fig1}}

\figcaption{Artificial spectral line with six clumps added.  The lower
spectrum indicates the result of the smoothing and subtraction
technique described in the text.  The six clumps are readily apparent,
along with the edges of the line, although the edges look distinctly
different from the clumps.  For this, and all subsequent figures of
smoothed and subtracted line profiles, the flux-density scaling is
linear but arbitrary; the profiles are offset by arbitrary, additive
amounts.\label{fig2}}

\figcaption{Smoothed and subtracted [\protect\ion{O}{1}]
$\lambda\lambda$6300, 6364 line on day 209.  Pairs of maxima and
minima with the same relative velocity for the [\protect\ion{O}{1}]
$\lambda$6300 and [\protect\ion{O}{1}] $\lambda$6364 lines are
indicated.  The zero point for velocity is 6300 \AA.  See discussion
in the text.\label{fig3}}

\figcaption{Time evolution of the [\protect\ion{O}{1}]
$\lambda\lambda$6300, 6364 line from day 93 to day 433.  The zero
point for velocity is 6300 \AA.  Note the remarkable consistency of
the substructure in the line.\label{fig4}}

\figcaption{Comparison of the clumps in the [\protect\ion{O}{1}]
$\lambda$5577, [\protect\ion{O}{1}] $\lambda\lambda$6300, 6364, and
\protect\ion{O}{1} $\lambda$7774 emission lines on day 209.  The zero
point for velocity for [\protect\ion{O}{1}] $\lambda\lambda$6300, 6364
is 6300 \AA.  All three have the same basic structure, although the
one permitted line ($\lambda$7774) is missing the component at $-2340$
km~s$^{-1}$; see text for details.\label{fig5}}

\figcaption{Time evolution of the \protect\ion{Mg}{1}] $\lambda$4571
line from day 139 to day 298.  The major clumps persist over time, but
there are changes in the details of the substructure.\label{fig6}}

\figcaption{Time evolution of the [\protect\ion{Ca}{2}]
$\lambda\lambda$7291, 7324 line from day 123 to day 298.  The zero
point for velocity is 7291 \AA.  The two components of the doublet
stand out over the entire series, with relatively little other
structure.\label{fig7}}

\figcaption{Comparison of the clumps in [\protect\ion{Ca}{2}]
$\lambda\lambda$7291, 7324, H$\alpha$, [\protect\ion{O}{1}]
$\lambda\lambda$6300, 6364, and \protect\ion{Mg}{1}] $\lambda$4571 on
day 209. The zero point for velocity for [\protect\ion{O}{1}]
$\lambda\lambda$6300, 6364 is 6300 \AA, while that for
[\protect\ion{Ca}{2}] $\lambda\lambda$7291, 7324 is 7291 \AA.  The
individual clumps in each species do not appear correlated.  The
magnesium line may show some clumps that line up with oxygen.  Calcium
and hydrogen are distinctly different.\label{fig8}}

\figcaption{Comparison of the clumps in [\protect\ion{O}{1}]
$\lambda\lambda$6300, 6364, H$\alpha$, and \protect\ion{Mg}{1}]
$\lambda$4571 on day 433.  The zero point for velocity for
[\protect\ion{O}{1}] $\lambda\lambda$6300, 6364 is 6300 \AA.  By late
times, there is no correspondence between the lines for the
substructure.  Note the narrow component of H$\alpha$ at zero
velocity.\label{fig9}}

\figcaption{Three examples of the late-time spectra of SN 1993J on
days 976, 1766, and 2454.  The day 976 spectrum was obtained under
non-photometric conditions, and so has been arbitrarily scaled to
match the other two.  A constant has also been added to the day 976
and day 1766 spectra to offset the spectra from each other.  The day
2454 spectrum suffers from minor second-order contamination beyond
$\sim$ 7600 \AA.  Note the relative increase over time of the oxygen
lines in comparison with H$\alpha$.  In addition, the substructure of
the H$\alpha$ line varies over time, although a small feature at zero
velocity persists.\label{fig10}}

\figcaption{Decomposition of the \protect\ion{He}{1} $\lambda$5876 +
\protect\ion{Na}{1}~D blend on day 976 using simple box-like profiles
to model the individual lines.  The dotted lines indicate the two
separate components (\protect\ion{He}{1} $\lambda$5876 and the
\protect\ion{Na}{1}~D blend), while the dashed line is the sum that
produces the overall blend.  This best fit indicates that the two
lines contribute approximately equally to the blend.\label{fig11}}

\figcaption{Comparison of the H$\alpha$ and H$\beta$ lines on days
1766 and 2454.  The red half of H$\beta$ is heavily contaminated by
[\protect\ion{O}{3}] $\lambda\lambda$4959, 5007, so it has been
removed for clarity.  The relatively strong, narrow H$\beta$ feature
at $\sim$ $-8,000$ km~s$^{-1}$ does not appear to correspond with the
structure of H$\alpha$, which is contaminated by [\protect\ion{O}{1}]
$\lambda\lambda$6300, 6364 on the blue side (cf. Figure
13).\label{fig12}}

\figcaption{Four oxygen lines on day 1766.  The two-horned structure
is clearly evident for [\protect\ion{O}{2}] $\lambda\lambda$7319, 7330
(zero point: 7325 \AA) and [\protect\ion{O}{3}] $\lambda\lambda$4959,
5007 (zero point: 5007 \AA).  The weakness of the lines, contamination
from other species, and the noisy nature of their regions of the
spectra obscure the features in [\protect\ion{O}{3}] $\lambda$4363 and
[\protect\ion{O}{1}] $\lambda\lambda$6300, 6364 (zero point: 6300
\AA), but they are still apparent.  The red peak of
[\protect\ion{O}{1}] $\lambda\lambda$6300, 6364 is visible as a
distinct shoulder on the blue side of H$\alpha$.  The consistency of
the two-horned structure over several lines, as well as the large
velocity separation of the two horns ($\sim$ 7100 km~s$^{-1}$),
indicates that these profiles do not represent the two separate
components of any of the oxygen doublets, but rather show the presence
of a flattened, disk-like structure.\label{fig13}}

\figcaption{Line profiles of [\protect\ion{O}{2}]
$\lambda\lambda$7319, 7330 on 976, 1766, and 2454.  The zero point for
the velocity is 7325 \AA.  The two-horned structure is obvious in the
last two spectra, but the beginnings are apparent even at day 976,
especially for the blue peak.\label{fig14}}

\figcaption{Spectra of three bright SNe illustrating the potential
impact of weak telluric absorption on high S/N ratio spectra.
Spectrum (a) is SN 1993J from day 19 (1993 April 15), (b) is SN 1994D
from 1994 April 18, (c) is SN 1999em from 1999 November 5, and (d) is
the same SN 1999em spectrum after an attempt to remove some of the
effects of telluric absorption.  All three SNe were observed with the
same instrument, but have different redshifts, indicating that the
feature at 6515 \AA\ is not intrinsic to the SNe.  When telluric
effects are considered, the feature can be removed reasonably well.
Note the presence of even weaker features (such as around 6470 \AA)
which we did not attempt to remove.\label{fig15}}

\figcaption{Spectrum of telluric absorption as determined from solar
Fourier-transform spectra from the National Solar Observatory database
at an airmass of 1.0.  The spectrum has been smoothed to 2 \AA/pixel,
and the relative strength of the water features is enhanced (see
text).  The inset shows the region around the narrow feature at 6515
\AA.\label{fig16}}

\clearpage

\begin{deluxetable}{lccccccc}

\tablenum{1}
\tabletypesize{\small}
\tablewidth{0pt}
\tablecaption{RELATIVE LINE FLUXES\label{tbl-1}}
\tablehead{\colhead{Day\tablenotemark{a}} &
\colhead{H$\alpha$} &
\colhead{\protect\ion{He}{1} + \protect\ion{Na}{1}}  &
\colhead{[\protect\ion{O}{1}]} &
\colhead{[\protect\ion{O}{2}]} &
\colhead{[\protect\ion{O}{3}] + H$\beta$} &
\colhead{[\protect\ion{O}{3}]} &
\colhead{[\protect\ion{Ne}{3}]}  \\
\colhead{} &
\colhead{$\lambda$6563} &
\colhead{$\lambda$5876} &
\colhead{$\lambda\lambda$6300,} &
\colhead{$\lambda\lambda$7319,} &
\colhead{$\lambda\lambda$4959,} &
\colhead{$\lambda$4363} &
\colhead{$\lambda$3967} \\
\colhead{} &
\colhead{} &
\colhead{$\lambda\lambda$5890,} &
\colhead{6364} &
\colhead{7330} &
\colhead{5007} &
\colhead{} &
\colhead{} \\
\colhead{} &
\colhead{} &
\colhead{5896} &
\colhead{} &
\colhead{} &
\colhead{$\lambda$4861} &
\colhead{} &
\colhead{}}
\startdata

553     & 100, 100& 30, 32&   8.8, 9.0&   23, 21  &  33, 39 &   34, 44&   45, 62   \\ 
670     & 100, 100& 29, 30&   4.9, 5.0&   23, 21  &  29, 35 &   27, 35&   \nodata \\ 
881     & 100, 100& 40, 43& 4.6, 4.6:  &   \nodata&  37, 44: & \nodata& \nodata   \\ 
976     & 100, 100& 37, 39&   3.2, 3.3&   25, 23  &  33, 40 &   27, 35&   \nodata \\ 
1766    & 100, 100& 54, 57&   6.8, 6.9&   46, 44  &  40, 48 &   19, 25&   \nodata \\ 
2028    & 100, 100& 82, 86:&   18, 18:  &   85, 80:  &  80, 94: &   33, 44:&   62, 85:   \\
2069    & 100, 100& 90, 95&  12, 12:   &   52, 50  &  87, 104&   27, 36:&   73, 100:  \\
2115    & 100, 100& 66, 70&  12, 12   &   68, 65  &  79, 94 &   28, 36:&   72, 100:  \\
2176    & 100, 100& 70, 74&  16, 16:   &   71, 67  &  67, 80 &  30, 39 &  96, 132:   \\ 
2454    & 100, 100& 74, 78&  13, 13   &   89, 84  &  76, 91 &  24, 31 & \nodata   \\
\enddata
\tablecomments{Listed relative to H$\alpha \equiv 100$.  First value is relative flux as 
directly measured; second value is relative flux corrected for a reddening of $A_V = 0.6$ mag using the extinction corrections of Cardelli et
al. (1989), including the O'Donnell (1994) modifications.  In all cases a
continuum was removed from beneath the individual lines.  When 
two (or more) lines are listed, the flux indicated is the sum for all.  Values are uncertain by $\sim$ 10\% (those
indicated with a colon are uncertain by $\gtrsim$ 20\%).}
\tablenotetext{a}{Day since explosion, 1993 March 27.5 UT.}
\end{deluxetable}
\clearpage
\begin{deluxetable}{lccccccc}

\tablenum{2}
\tabletypesize{\small}
\tablewidth{0pt}
\tablecaption{MEASURED VELOCITIES\label{tbl-2}}
\tablehead{\colhead{Day\tablenotemark{a}} &
\colhead{H$\alpha$ $\lambda$6563} &
\colhead{H$\alpha$ $\lambda$6563} &
\colhead{H$\alpha$ $\lambda$6563} &
\colhead{H$\beta\ \lambda$4861} &
\colhead{[\protect\ion{O}{3}] $\lambda$5007} &
\colhead{\protect\ion{He}{1} $\lambda$5876}  &
\colhead{\protect\ion{Na}{1} $\lambda$5890} \\
\colhead{} &
\colhead{BVZI\tablenotemark{b}} & 
\colhead{RVZI\tablenotemark{b}} & 
\colhead{FWHM\tablenotemark{b}} & 
\colhead{BVZI\tablenotemark{b}} & 
\colhead{RVZI\tablenotemark{b}} & 
\colhead{BVZI\tablenotemark{b}} & 
\colhead{RVZI\tablenotemark{b}} } 
\startdata

433     &  $-$16600 &  11700 &   23100& \nodata & \nodata  & \nodata & \nodata \\
473     &  $-$16100 &  10900 &   22000& \nodata & \nodata  & \nodata & \nodata \\
523     &  $-$10400 &  13100:&   17400& \nodata & \nodata  & \nodata & \nodata \\
553     &  $-$10000 &  9800  &   17200&  $-$10500 &   9100   &  $-$16200 &   8400  \\
670     &  $-$10500 &  10100 &   16900&   $-$9700 &    9600  &  $-$16000 &    9200 \\
881     &  $-$10300 &  9500  &   16400&   $-$9500 &  10000   &  $-$16200 &   8700  \\
976     &  $-$10600 &   9800 &   16000&   $-$9500 &    8000  &  $-$16000 &    8900 \\
1766    &  $-$9900  &   9700 &   15000&   $-$9700 &    7600: &  $-$15600 &    8500   \\
2028    &  $-$8900  &   9100 &   15000&  $-$10400 &    9800  &  $-$21100:&    8400  \\
2069    &  $-$9400  &   9100 &   15000&  $-$12000 &    7500  &  $-$17200 &    8700  \\
2115    &  $-$9900  &   9300 &   14700&  $-$14900:&    7200: &  $-$15400 &    8300   \\
2176    & $-$9600   &   9000 &   14900&  $-$11800:&    6700: &  $-$15900 &    7700   \\
2454    & $-$9300   &   8700 &   13800&  $-$13400:&    8400: &  $-$15200 &    7500   \\

\enddata
\tablecomments{All velocities are km~s$^{-1}$.  Values are uncertain by $\sim$ 10\% (those indicated by a colon are
uncertain by $\sim 20-30$\%).}.
\tablenotetext{a}{Day since explosion, 1993 March 27.5 UT.}
\tablenotetext{b}{BVZI = blue velocity at zero intensity; RVZI = red velocity at zero intensity; FWHM = full width at
half maximum.}

\end{deluxetable}

\clearpage

\begin{deluxetable}{lccc}

\tablenum{3}

\tablewidth{300pt}
\tablecaption{LINE DECOMPOSITION FLUXES\label{tbl-3}}
\tablehead{\colhead{Day\tablenotemark{a}} &
\colhead{H$\beta$\tablenotemark{b}} &
\colhead{[\protect\ion{O}{1}]\tablenotemark{c}} &
\colhead{[\protect\ion{O}{3}]\tablenotemark{c}}  \\
\colhead{} &
\colhead{$\lambda$6563} &
\colhead{$\lambda$6300} &
\colhead{$\lambda$5007} }

\startdata
976  &  0.2, 0.3  &  0.3, 0.3  &  1.5, 1.9  \\
1766 &  0.2, 0.2  &  0.3, 0.3  &  1.3, 1.6  \\
2028 &  0.3, 0.4  &  0.3, 0.4  &  1.3, 1.8  \\
2069 &  0.3, 0.4  &  0.4, 0.4  &  1.9, 2.3  \\
2115 &  0.2, 0.3  &  0.3, 0.3  &  1.5, 1.8  \\
2176 &  0.2, 0.3  &  0.3, 0.4  &  1.4, 1.7  \\
2454 &  0.1, 0.1  &  0.3, 0.3  &  1.2, 1.4  \\
\enddata
\tablecomments{Line fluxes as a fraction of another line in the spectrum.  Details of decomposition are presented in the
text.  The first number is the fraction as measured; the second is the fraction after the spectrum has been dereddened
by $A_V = 0.6$ mag using extinction corrections of Cardelli et
al. (1989), including the O'Donnell (1994) modifications.  The values for days 2028, 2069, 2115, and 2176 are much less
certain than those for days 976, 1766, and 2454.}
\tablenotetext{a}{Day since explosion, 1993 March 27.5 UT.}
\tablenotetext{b}{Fraction relative to the strength of H$\alpha$.}
\tablenotetext{c}{Fraction relative to the strength of \protect[\ion{O}{2}] $\lambda\lambda$7319, 7323.}

\end{deluxetable}
\clearpage

\begin{deluxetable}{lccccccc}

\tablenum{4}
\tabletypesize{\small}
\tablewidth{0pt}
\tablecaption{PREDICTED LINE FLUXES\label{tbl-4}}
\tablehead{\colhead{Age} &
\colhead{H$\alpha$} &
\colhead{H$\beta$} &
\colhead{[\protect\ion{O}{1}]} &
\colhead{[\protect\ion{O}{2}]} &
\colhead{[\protect\ion{O}{3}]} &
\colhead{[\protect\ion{Ne}{3}]} &
\colhead{\protect\ion{Na}{1}~D} \\
\colhead{(yr)} &
\colhead{$\lambda$6563} &
\colhead{$\lambda$4861} &
\colhead{$\lambda\lambda$6300,} &
\colhead{$\lambda\lambda$7319,} &
\colhead{$\lambda\lambda$4959,} &
\colhead{$\lambda\lambda$3868,} &
\colhead{$\lambda\lambda$5890,} \\
\colhead{} &
\colhead{} &
\colhead{} &
\colhead{6364} &
\colhead{7330} &
\colhead{5007} &
\colhead{3969} &
\colhead{5896} }
\startdata
PL\tablenotemark{a} &  &     &    &    &     &    &     \\
 \ \  1     & 100 & 14 & 26 &  3 &   7 & 11 &  27 \\
 \ \  2     & 100 & 23 & 26 &  6 &  27 & 18 &  38 \\
  \ \ 5     & 100 & 30 & 35 & 14 & 100 & 30 &  61 \\
  \ \ 10    & 100 & 32 & 52 & 15 & 220 & 43 & 100 \\
  \ \ 17.5  & 100 & 34 & 72 & 14 & 470 & 55 & 140 \\
  \ \ 30    & 100 & 34 & 74 & 12 & 640 & 56 & 140 \\
 RSG\tablenotemark{b} &  &     &    &    &     &    &     \\
  \ \ 10    & 100 & 33 & 14 &\nodata & 340 & 30 & 17 \\
\enddata
\tablecomments{Line fluxes from the circumstellar interaction model of Chevalier \& Fransson (1994), listed relative to
H$\alpha \equiv 100$.  Only the lines relevant to our observations are shown.}
\tablenotetext{a}{Power-law model.}

\tablenotetext{b}{Red-supergiant model.}

\end{deluxetable}
\clearpage



\begin{thebibliography}{}
\bibitem{} Aldering, G., Humphreys, R. M., \& Richmond, M. 1994, \aj,
107, 662
\bibitem{} Almog, Y., \& Netzer, H. 1989, \mnras, 238, 57
\bibitem{} Anderson, M. C., Jones, T. W., Rudnick, L., Tregillis,
I. L., \& Kang, H. 1994, \apj, 421, L31
\bibitem{} Andronova, A. A. 1992, Sov. Astron. Lett., 18, 360
\bibitem{} Aretxaga, I., Benetti, S., Terlevich, R. J.,
 Fabian, A. C., Cappellaro, E., Turatto, M., \&
 Della Valle, M. 1999, \mnras, 309, 343
\bibitem{} Aschenbach, B., Egger, R., \& Tr\"umper, J. 1995, \nat, 373,
587
\bibitem{} Baade, W., \& Minkowski, R. 1954, \apj, 119, 206
\bibitem{} Barbon, R., Benetti, S., Cappellaro, E., Patat, F.,
Turatto, M., \& Iijima, T. 1995, \aaps, 110, 513
\bibitem{} Benetti, S., Cappellaro, E., Danziger, I. J., Turatto, M.,
Patat, F., \& Della Valle, M. 1998, \mnras, 294, 448
\bibitem{} Bessell, M. S. 1999, \pasp, 111, 1426
\bibitem{} Blondin, J. M., Lundqvist, P., \& Chevalier, R. A. 1996,
\apj, 472, 257
\bibitem{} Bouchet, P., Lawrence, S., Crotts, A., Sugerman, B.,
Uglesich, R., \& Heathcote, S. 2000, \iaucirc\ 7354
\bibitem{} Branch, D., Livio, M., Yungelson, L. R., Boffi, F. R., \&
Baron, E. 1995, \pasp, 107, 1019
\bibitem{} Cappellaro, E., Danziger, I. J., \& Turatto, M. 1995,
\mnras, 277, 106
\bibitem{} Cardelli, J. A., Clayton, G. C., \& Mathis, J. S. 1989,
\apj, 345, 245
\bibitem{} Chevalier, R. A., \& Blondin, J. M. 1995, \apj, 444, 312
\bibitem{} Chevalier, R. A., Blondin, J. M., \& Emmering, R. T. 1992,
\apj, 392, 118
\bibitem{} Chevalier, R. A., \& Fransson, C. 1989, \apj, 343, 323
\bibitem{} Chevalier, R. A., \& Fransson, C. 1992, \apj, 395, 540
\bibitem{} Chevalier, R. A., \& Fransson, C. 1994, \apj, 420, 268 (CF94)
\bibitem{} Chu, Y.-H., Caulet, A., Montes, M., Panagia, N., Van Dyk,
S. D., \& Weiler, K. W. 1999, \apj, 512, L51
\bibitem{} Chugai, N. N. 1992, Sov. Astron. Lett., 18, 239
\bibitem{} Chugai, N. N. 1994, \apj, 428, L17
\bibitem{} Chugai, N. N. 1986, Sov. Astron., 30, 563
\bibitem{} Chugai, N. N., Danziger, I. J., \& Della Valle, M. 1995,
\mnras, 276, 530
\bibitem{} Clocchiatti, A., Wheeler, J. C., Barker, E. S., Filippenko,
A. V., Matheson, T., \& Liebert, J. W. 1995, \apj, 446, 167
\bibitem{} Fassia, A., Meikle, W. P. S., Geballe, T. R., Walton,
N. A., Pollacco, D. L., Rutten, R. G. M., \& Tinney, C. 1998, \mnras,
299, 150
\bibitem{} Fesen, R. A. 1993, \apj, 413, L109
\bibitem{} Fesen, R. A. 1998, \aj, 115, 1107
\bibitem{} Fesen, R. A., \& Becker, R. H. 1990, \apj, 351, 437
\bibitem{} Fesen, R. A., et al. 1999, \aj, 117, 725 (F99)
\bibitem{} Fesen, R. A., \& Matonick, D. M. 1993, \apj, 407, 110
\bibitem{} Fesen, R. A., \& Matonick, D. M. 1994, \apj, 428, 157
\bibitem{} Filippenko, A. V. 1982, \pasp, 94, 715
\bibitem{} Filippenko, A. V. 1989, \aj, 97, 726
\bibitem{} Filippenko, A. V. 1991a, in Supernovae, ed. S. E. Woosley
(New York: Springer-Verlag), 467
\bibitem{} Filippenko, A. V. 1991b, in SN 1987A and Other Supernovae,
eds. I. J. Danziger \& K. Kj\"ar (Garching: ESO), 343
\bibitem{} Filippenko, A. V., \& Sargent, W. L. W. 1989, \apj, 345, L43
\bibitem{} Filippenko, A. V., Matheson, T., \& Barth, A. J. 1994, \aj,
108, 2220 (FMB94)
\bibitem{} Filippenko, A. V., Matheson, T., \& Ho, L. C. 1993, \apj,
415, L103 (FMH93)
\bibitem{} Finn, R. A., Fesen, R. A., Darling, G. W., Thorstensen,
J. R., \& Worthey, G. S. 1995, \aj, 110, 300
\bibitem{} Fransson, C. 1984, \aap, 133, 264
\bibitem{} Fransson, C., Lundqvist, P., \& Chevalier, R. A. 1996,
\apj, 461, 993
\bibitem{} Freedman, W. L., et al. 1994, \apj, 427, 628
\bibitem{} Gerardy, C. L., Fesen, R. A., H\"oflich, P., \& Wheeler,
J. C. 2000, \apj, submitted (astro-ph/9912433)
\bibitem{} Haas, M. R., Colgan, S. W. J., Erickson, E. F., Lord,
S. D., Burton, M. G., \& Hollenbach, D. J. 1990, \apj, 360, 257
\bibitem{} Hachisu, I., Matsuda, T., Nomoto, K., \& Shigeyama,
T. 1991, \apj, 368, L27
\bibitem{} Hanuschik, R. W., Spyromilio, J., Stathakis, R.,
Kimeswenger, S., Gochermann, J., Seidensticker, K. J., \& Meurer,
G. 1993, 261, 909
\bibitem{} Houck, J. C., \& Fransson, C. 1996, \apj, 456, 811
\bibitem{} Iwamoto, K., Young, T. R., Nakasato, N., Shigeyama, T.,
Nomoto, K., Hachisu, I., \& Saio, H. 1997, \apj, 477, 865
\bibitem{} Kingdon, J., Ferland, G. J., \& Feibelman, W. A. 1995,
\apj, 439, 793
\bibitem{} Khokhlov, A. M., H\"oflich, P. A., Oran, E. S., Wheeler,
J. C., Wang, L., \& Chtchelkanova, A. Yu. 1999, \apj, 524, L107
\bibitem{} Kifonidis, K., Plewa, T., Janka, H.-Th., \& M\"uller,
E. 2000, \apj, submitted (astro-ph/9911183)
\bibitem{} Kondratyev, K. Ya. 1969, Radiation in the Atmosphere (New
York: Academic Press)
\bibitem{} Leibundgut, B., Kirshner, R. P., Pinto, P. A., Rupen,
M. P., Smith, R. C., Gunn, J. E., \& Schneider, D. P. 1991, \apj, 372,
531
\bibitem{} Leonard, D. C., Filippenko, A. V., Barth, A. J., \&
Matheson, T. 2000, \apj, in press (astro-ph/9908040)
\bibitem{} Lewis, J. R., et al. 1994, \mnras, 266, L27
\bibitem{} Li, A., Hu, J., Wang, L., Jiang, X., \& Li, H. 1994, \apss,
211, 323
\bibitem{} Li, H., \& McCray, R. 1992, \apj, 387, 309
\bibitem{} Li, H., \& McCray, R. 1993, \apj, 405, 730
\bibitem{} Livne, E., Tuchman, Y., \& Wheeler, J. C. 1992, \apj, 399,
665
\bibitem{} Long, K. S., Blair, W. P., \& Krzeminski, W. 1989, \apj,
340, L25
\bibitem{} Marcaide, J. M., et al. 1997, \apj, 486, L31
\bibitem{} Marietta, E., Burrows, A., \& Fryxell, B. 2000, \apj,
submitted (astro-ph/9908116)
\bibitem{} Matheson, T., et al. 2000, \aj, in press (Paper I)
\bibitem{} Nomoto, K., Suzuki, T., Shigeyama, T., Kumagai, S.,
Yamaoka, H., \& Saio, H. 1993, \nat, 364, 507
\bibitem{} O'Donnell, J. E. 1994, \apj, 422, 158
\bibitem{} Panagia, N., Scuderi, S., Gilmozzi, R., Challis, P. M.,
Garnavich, P. M., \& Kirshner, R. P. 1996, \apj, 459, L17
\bibitem{} Patat, F., Chugai, H., \& Mazzali, P. A. 1995, \aap, 299,
715
\bibitem{} Richmond, M. W., Treffers, R. R., Filippenko, A. V., Paik,
Y., Leibundgut, B., Schulman, E., \& Cox, C. V. 1994, \aj, 107, 1022
\bibitem{} Richmond, M. W., Treffers, R. R., Filippenko, A. V., \&
Paik, Y. 1996, \aj, 112, 732
\bibitem{} Ryder, S., Staveley-Smith, L., Dopita, M., Petre, R.,
Colbert, E., Malin, D., \& Schlegel, E. 1993, \apj, 416, 167
\bibitem{} Schachter, J. 1991, \pasp, 103, 457
\bibitem{} Schachter, J., Filippenko, A. V., \& Kahn, S. M. 1989,
\apj, 340, 1049
\bibitem{} Schlegel, D. J., Finkbeiner, D. P., \& Davis, M. 1998,
\apj, 500, 525
\bibitem{} Schlegel, E. M. 1990, \mnras, 244, 269
\bibitem{} Schlegel, E. M., \& Kirshner, R. P. 1989, \aj, 98, 577
\bibitem{} Shigeyama, T., Nomoto, K., Tsujimoto, T., \& Hashimoto,
M. 1990, \apj, 361, L23
\bibitem{} Spyromilio, J. 1991, \mnras, 253, 25P
\bibitem{} Spyromilio, J. 1994, \mnras, 266, L61 (S94)
\bibitem{} Spyromilio, J., Meikle, W. P. S., \& Allen, D. A. 1990,
\mnras, 242, 669
\bibitem{} Spyromilio, J., \& Pinto, P. A. 1991, in SN 1987A and Other
Supernovae, eds. I. J. Danziger \& K. Kj\"ar (Garching: ESO), 423
\bibitem{} Spyromilio, J., Stathakis, R. A., \& Meurer, G. R. 1993,
\mnras, 263, 530
\bibitem{} Stathakis, R. A., Dopita, M. A., Cannon, R. D., \& Sadler,
E. M. 1991, in Supernovae, ed. S. E. Woosley (New York: Springer), 95
\bibitem{} Stathakis, R. A., \& Sadler, E. M. 1991, \mnras, 250, 786
\bibitem{} Strom, R., Johnston, H. M., Verbunt, F., \& Aschenbach,
B. 1995, \nat, 373, 590
\bibitem{} Swartz, D. A. 1991, in Supernovae, ed. S. E. Woosley (New
York: Springer), 434
\bibitem{} Swartz, D. A., Clocchiatti, A., Benjamin, R., Lester,
D. F., \& Wheeler, J. C. 1993, \nat, 365, 232
\bibitem{} Swartz, D. A., Harkness, R. P., \& Wheeler, J. C. 1989,
\nat, 337, 439
\bibitem{} Trammell, S. R., Hines, D. C., \& Wheeler, J. C. 1993,
\apj, 414, L21
\bibitem{} Tran, H. D., Filippenko, A. V., Schmidt, G. D., Bjorkman,
K. S., Jannuzi, B. T., \& Smith, P. S. 1997, \pasp, 109, 489
\bibitem{} Turatto, M., Cappellaro, E., Danziger, I. J., Benetti, S.,
Gouiffes, C., \& Della Valle, M. 1993, \mnras, 262, 128 
\bibitem{} Uomoto, A. 1991, \aj, 101, 1275
\bibitem{} Utrobin, V. 1994, \aap, 281, L89
\bibitem{} Van Dyk, S. D., Weiler, K. W., Sramek, R. A., Rupen, M. P.,
\& Panagia, N. 1994, \apj, 324, L115
\bibitem{} Wade, R. A., \& Horne, K. D. 1988, \apj, 324, 411
\bibitem{} Wallace, L., Hinkle, K., \& Livingston, W. 1993, An Atlas of
the Photospheric Spectrum from 8900 to 13,600 cm$^{-1}$ (7350 to 11,230
\AA) (NSO: Tech. Rep. 93-001)
\bibitem{} Wallace, L., Hinkle, K., \& Livingston, W. 1998, An Atlas of the 
Spectrum of the Solar Photosphere from 13,500 to 28,000 cm$^{-1}$ (3570 to 
7405 \AA) (NSO: Tech. Rep. 98-001)
\bibitem{} Wang, L., Howell, D. A., H\"oflich, P., \& Wheeler,
J. C. 2000, \apj, submitted (astro-ph/9912033)
\bibitem{} Wang, L., \& Hu, J. 1994, \nat, 369, 380
\bibitem{} Wheeler, J. C., et al. 1993, \apj, 417, L71
\bibitem{} Wheeler, J. C., \& Filippenko, A. V. 1996, in Supernovae and
Supernova Remnants, ed. R. A. McCray \& Z. Wang (Cambridge: Cambridge
University Press), 241
\bibitem{} Wheeler, J. C., Harkness, R. P., Barker, E. S., Cochran,
A. L., \& Wills, D. 1987, \apj, 313, L69
\bibitem{} Wheeler, J. C., Swartz, D. A., \& Harkness, R. P. 1993,
\physrep, 227, 113
\bibitem{} Winkler, P. F., \& Kirshner, R. P. 1985, \apj, 299, 981
\bibitem{} Woosley, S. E., Eastman, R. G., Weaver, T. A., \& Pinto,
P. A. 1994, \apj, 429, 300
\bibitem{} Woosley, S. E., Langer, N., \&  Weaver, T. A. 1995, \apj,
448, 315

\end{thebibliography}
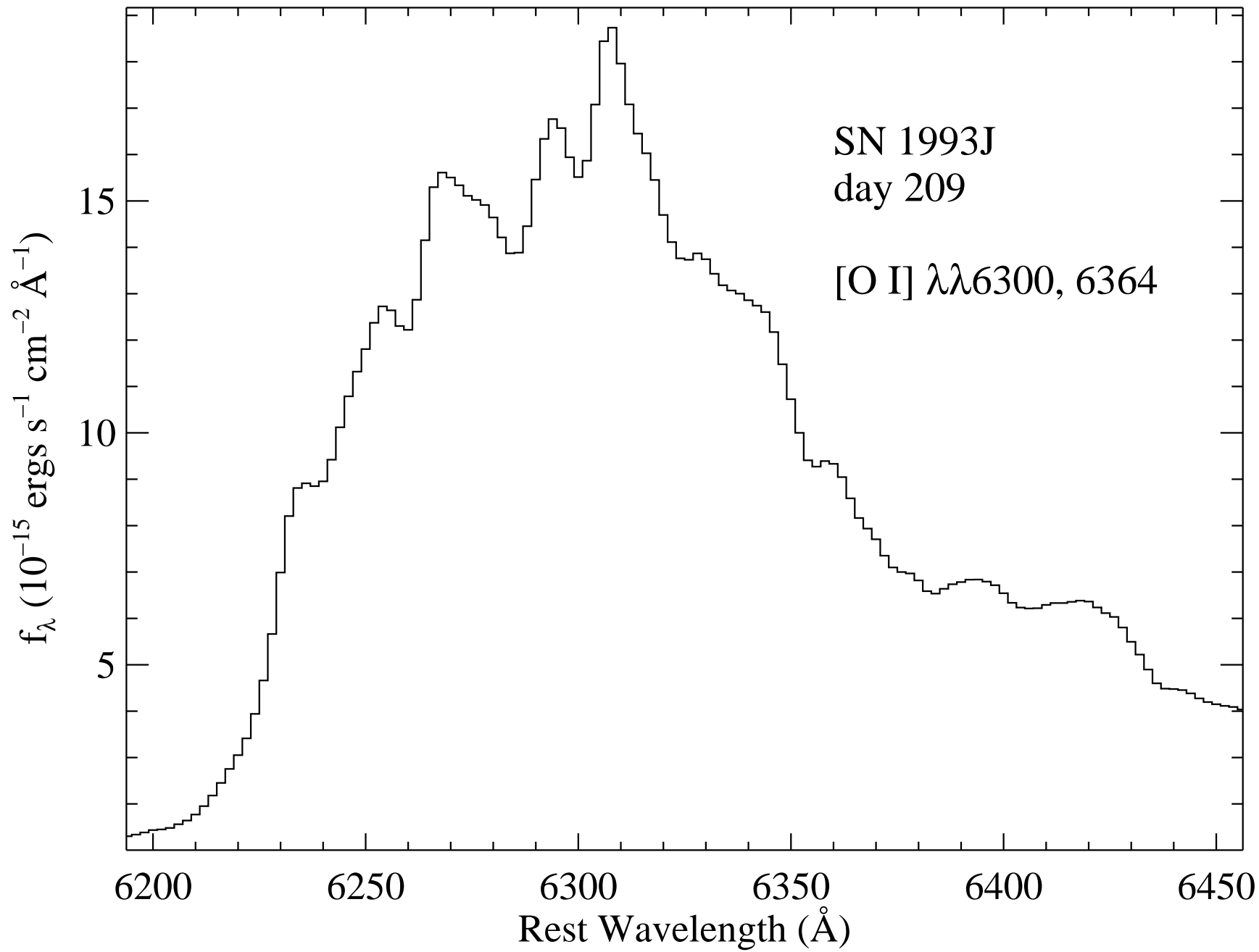
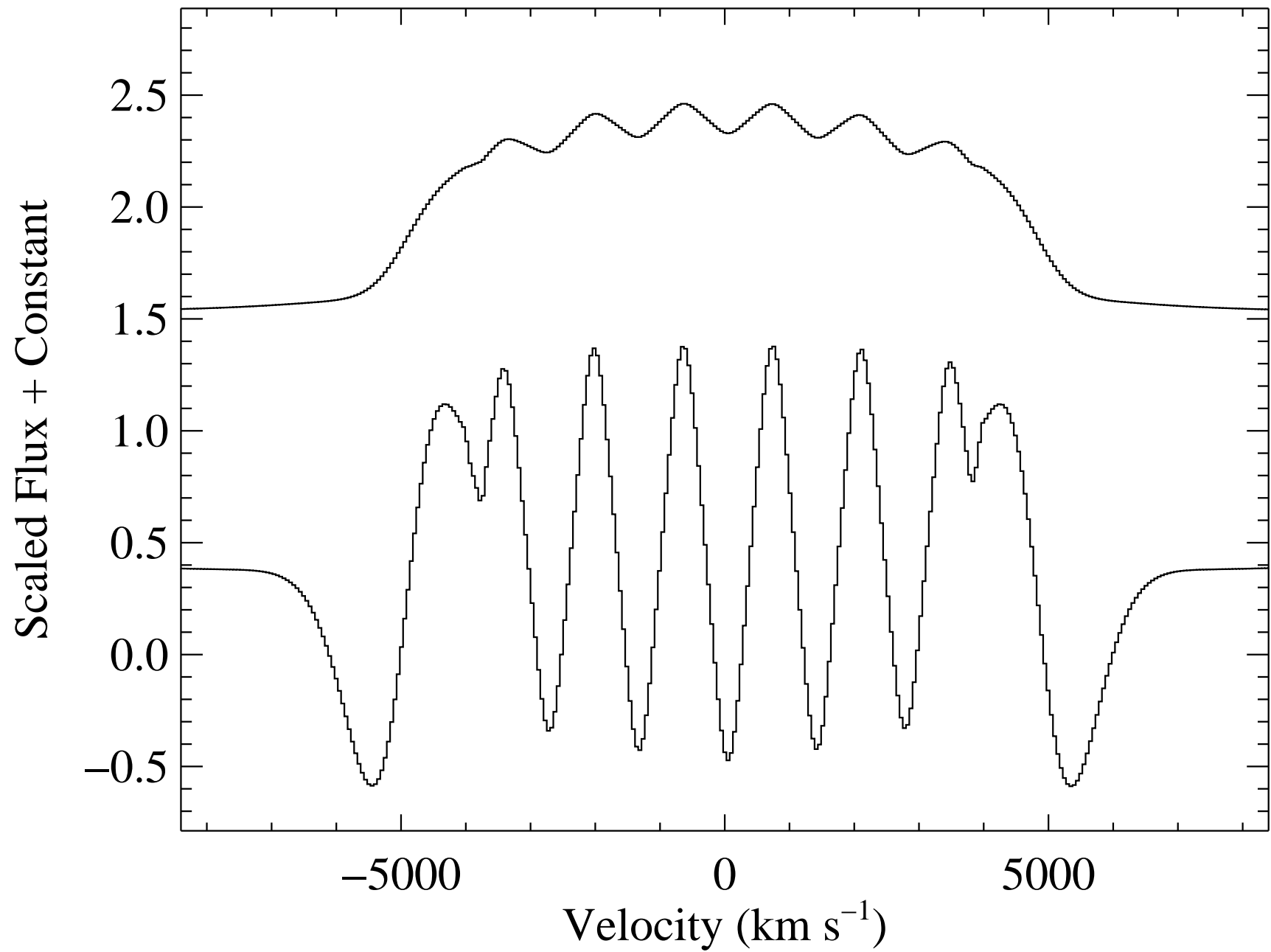
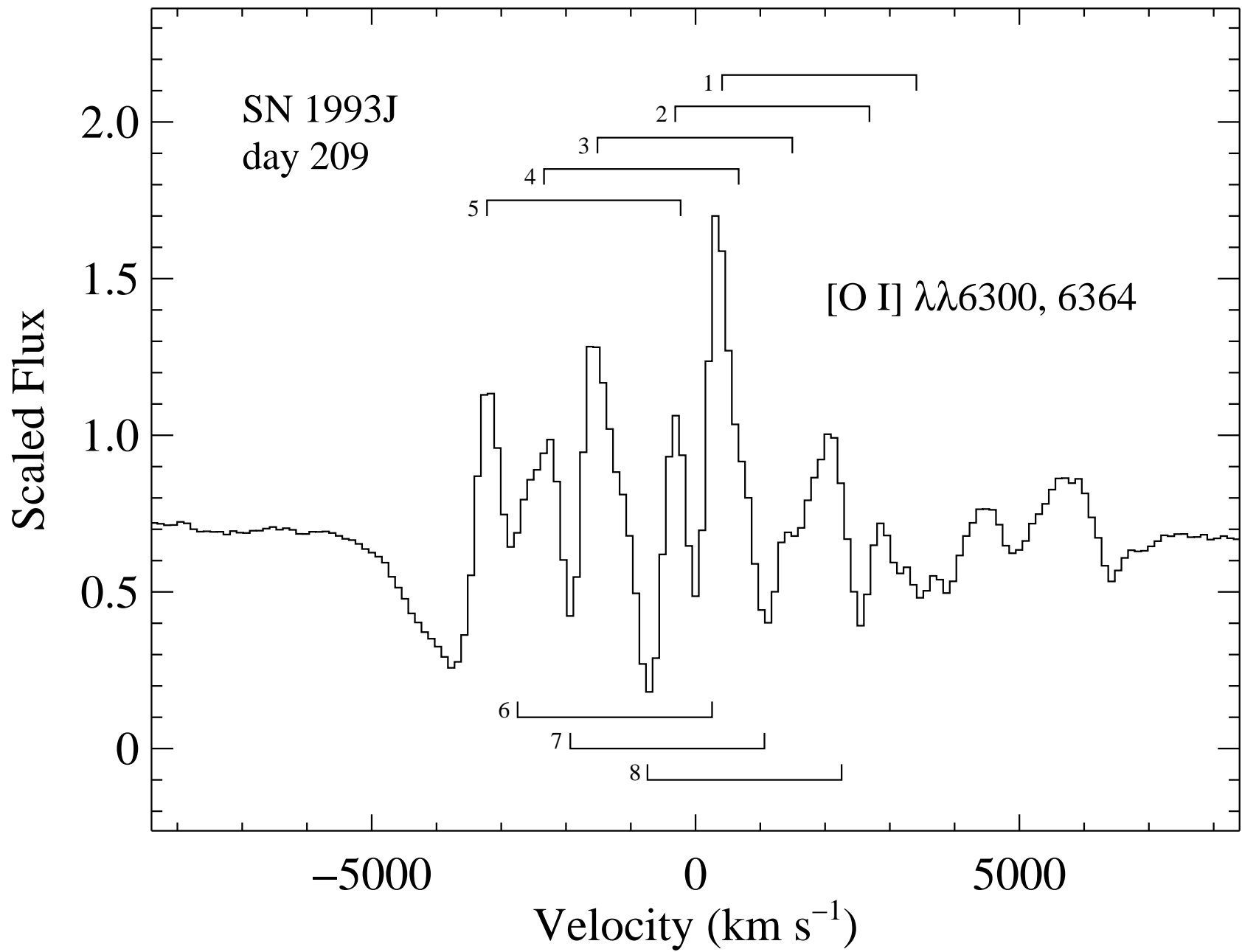
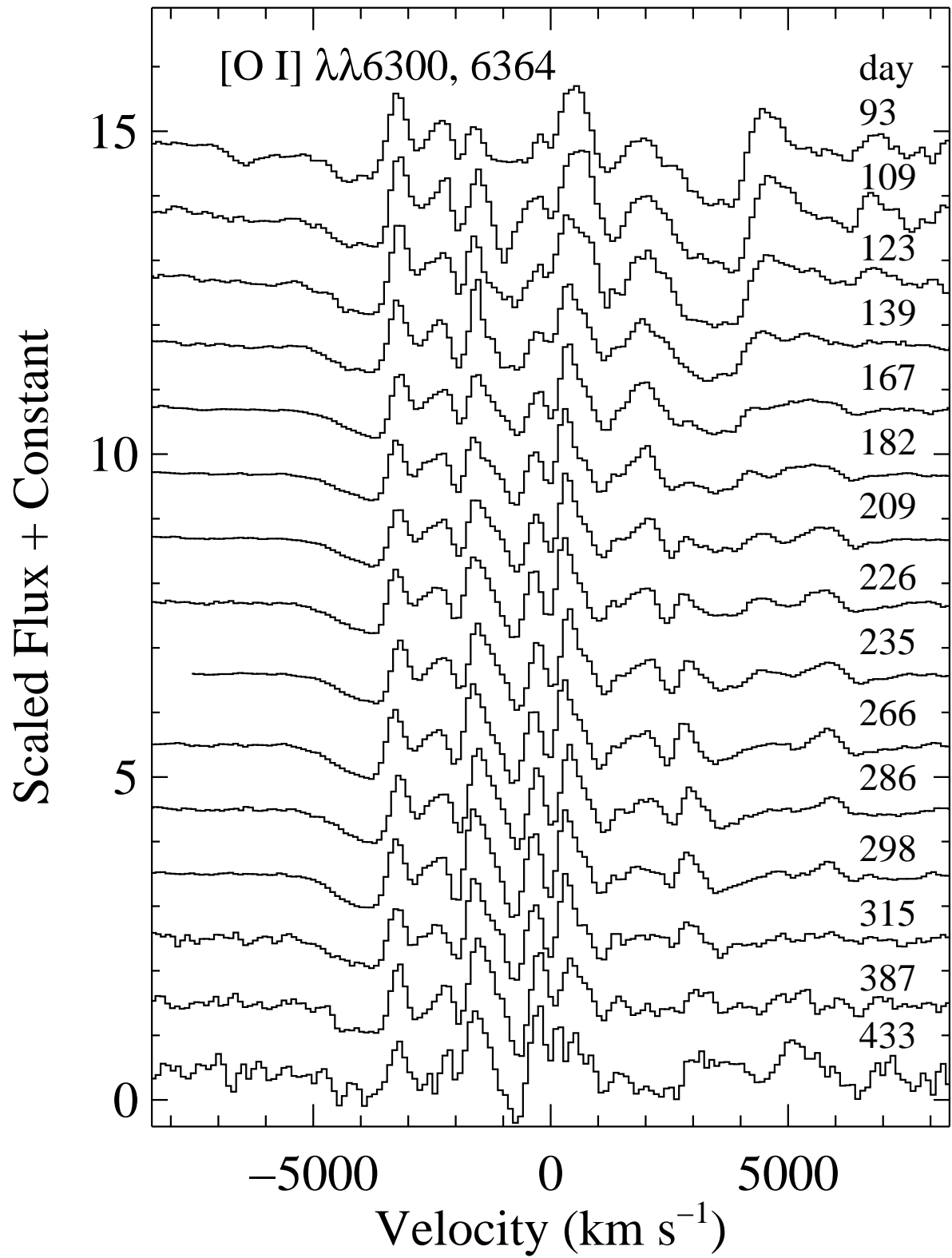
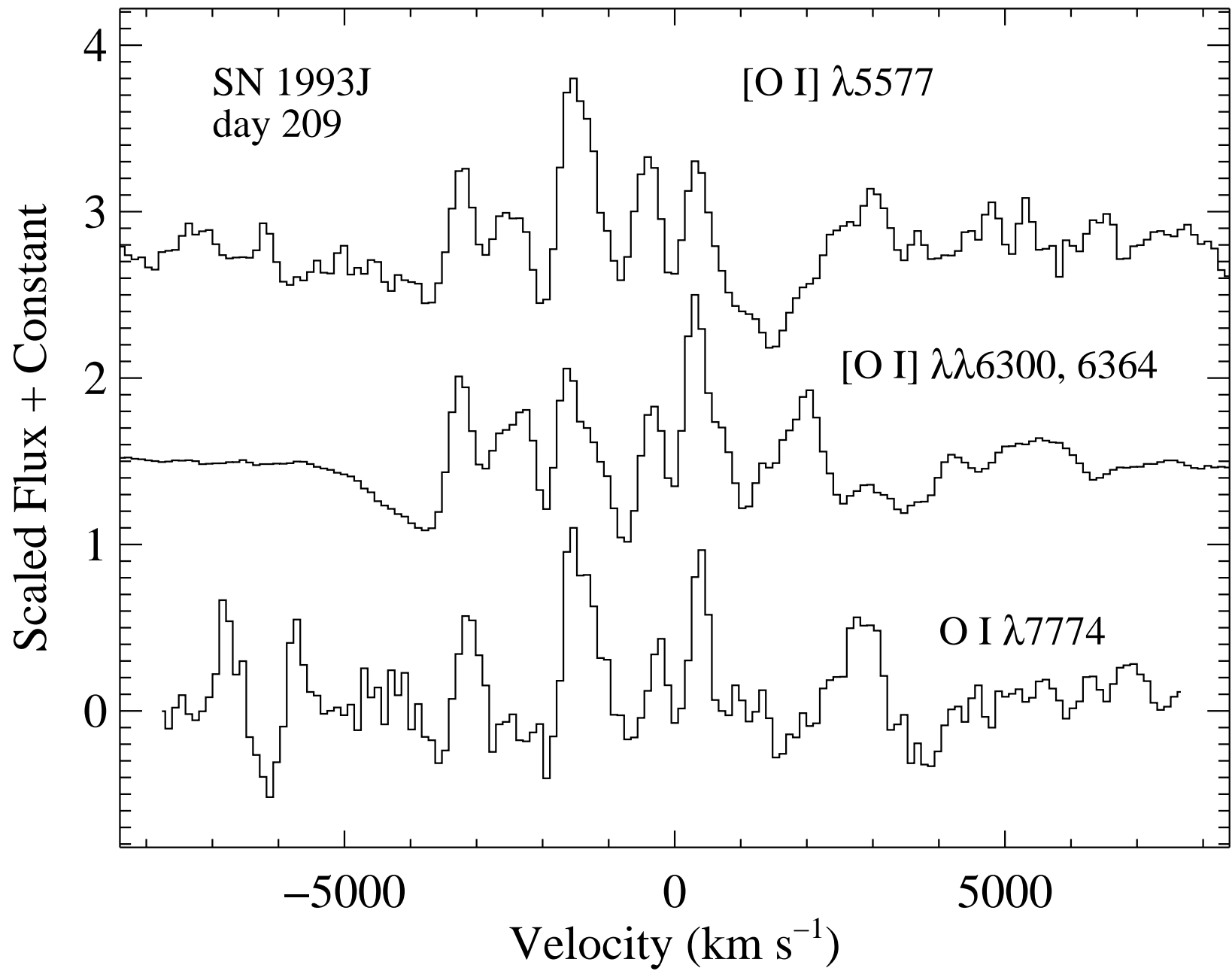
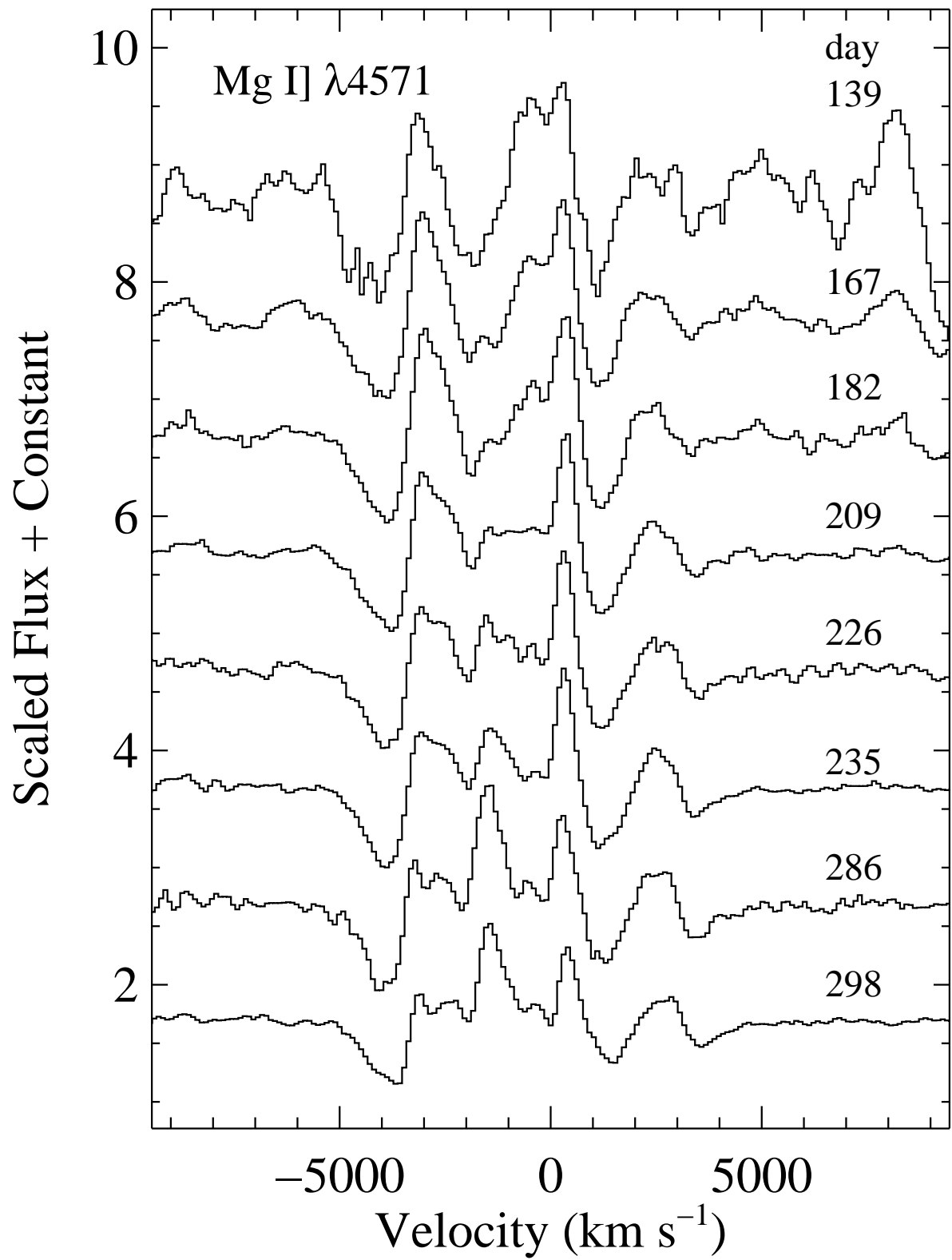
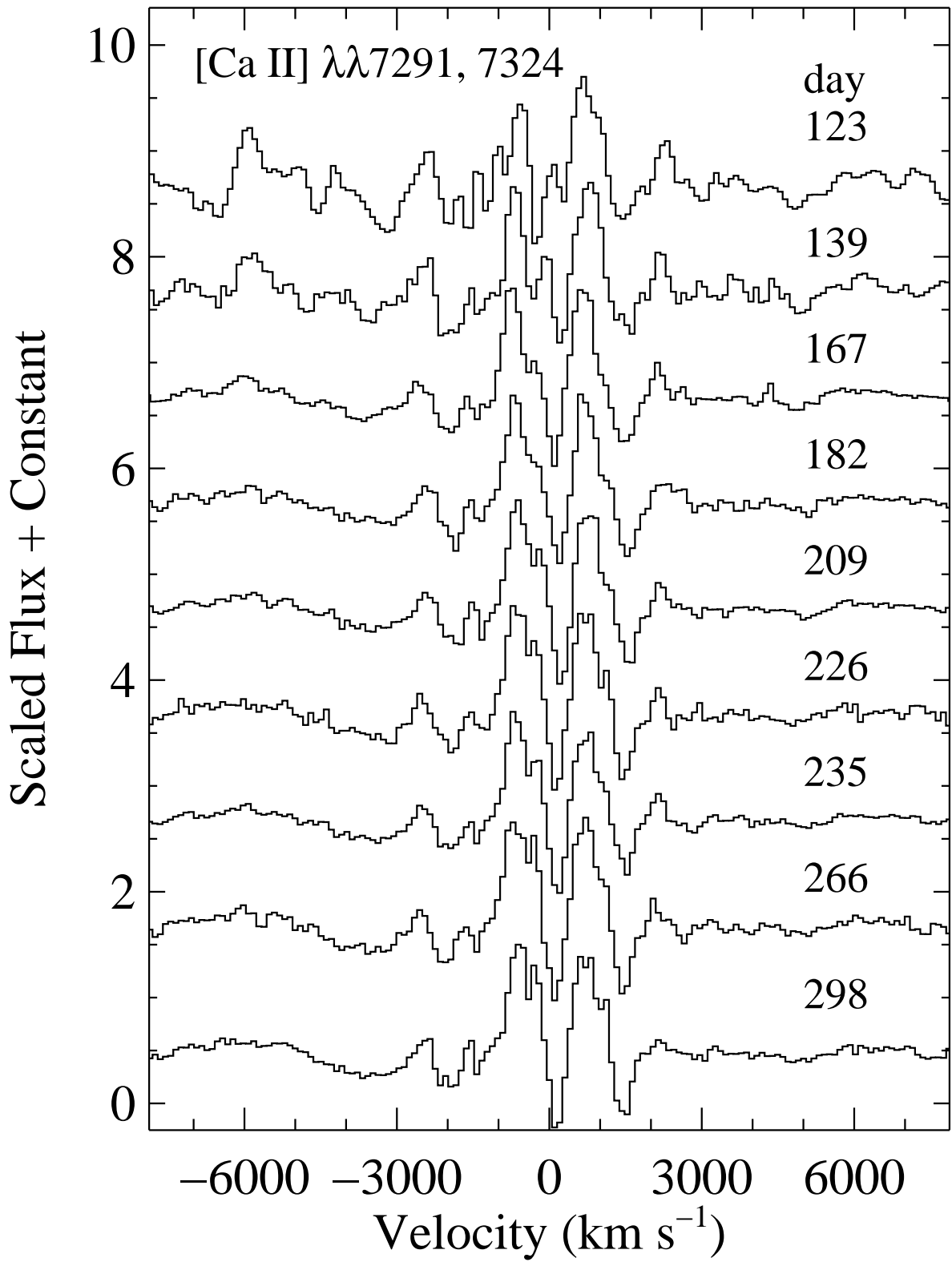
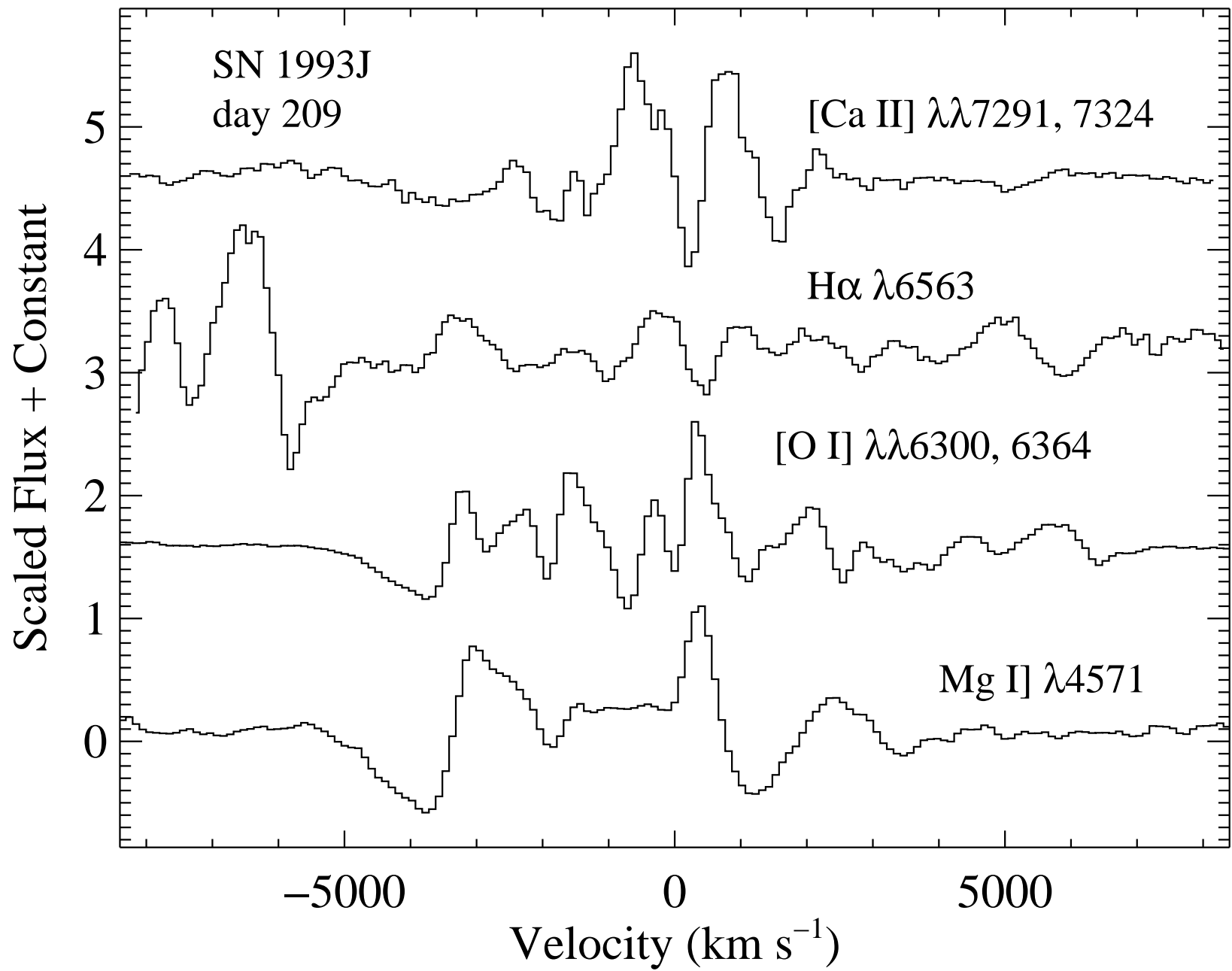
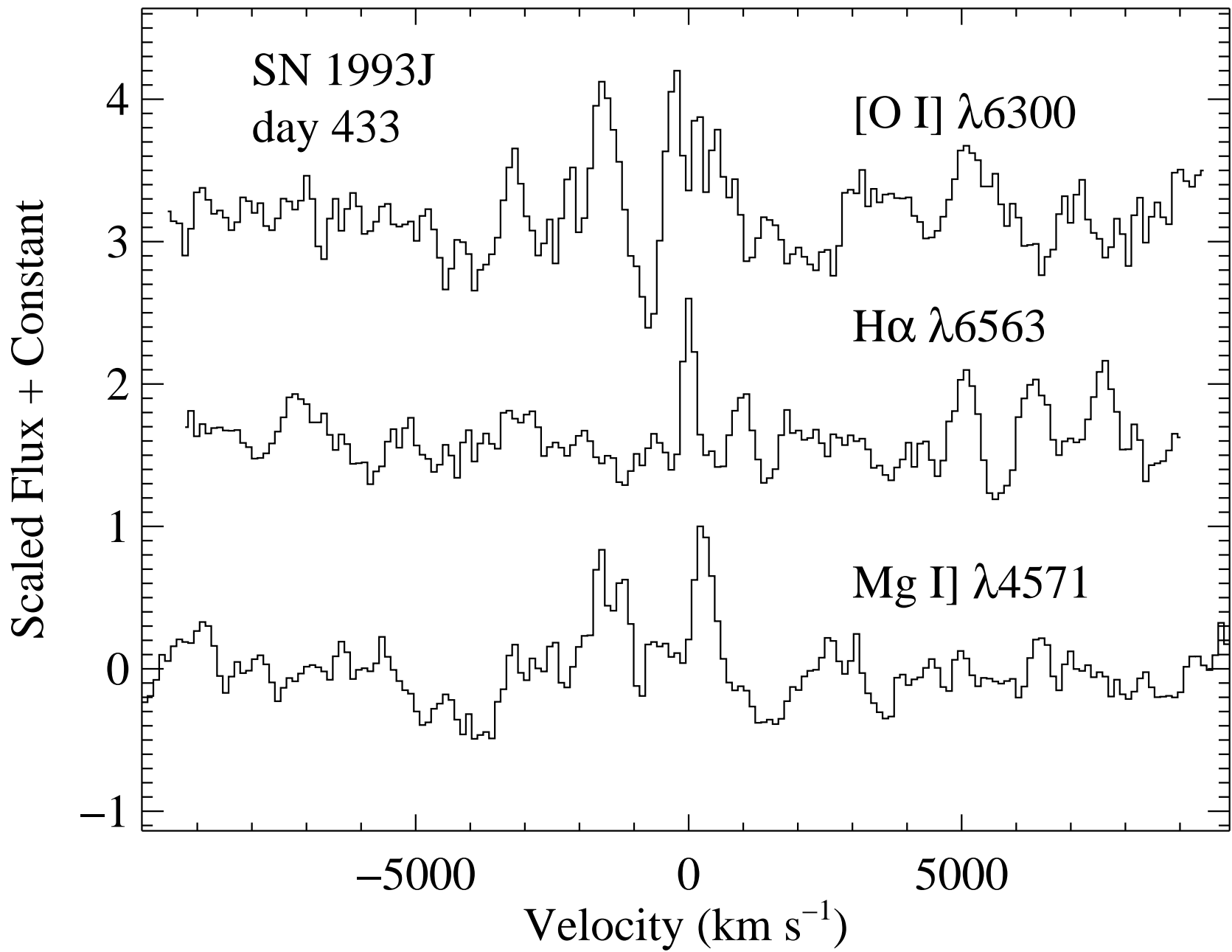
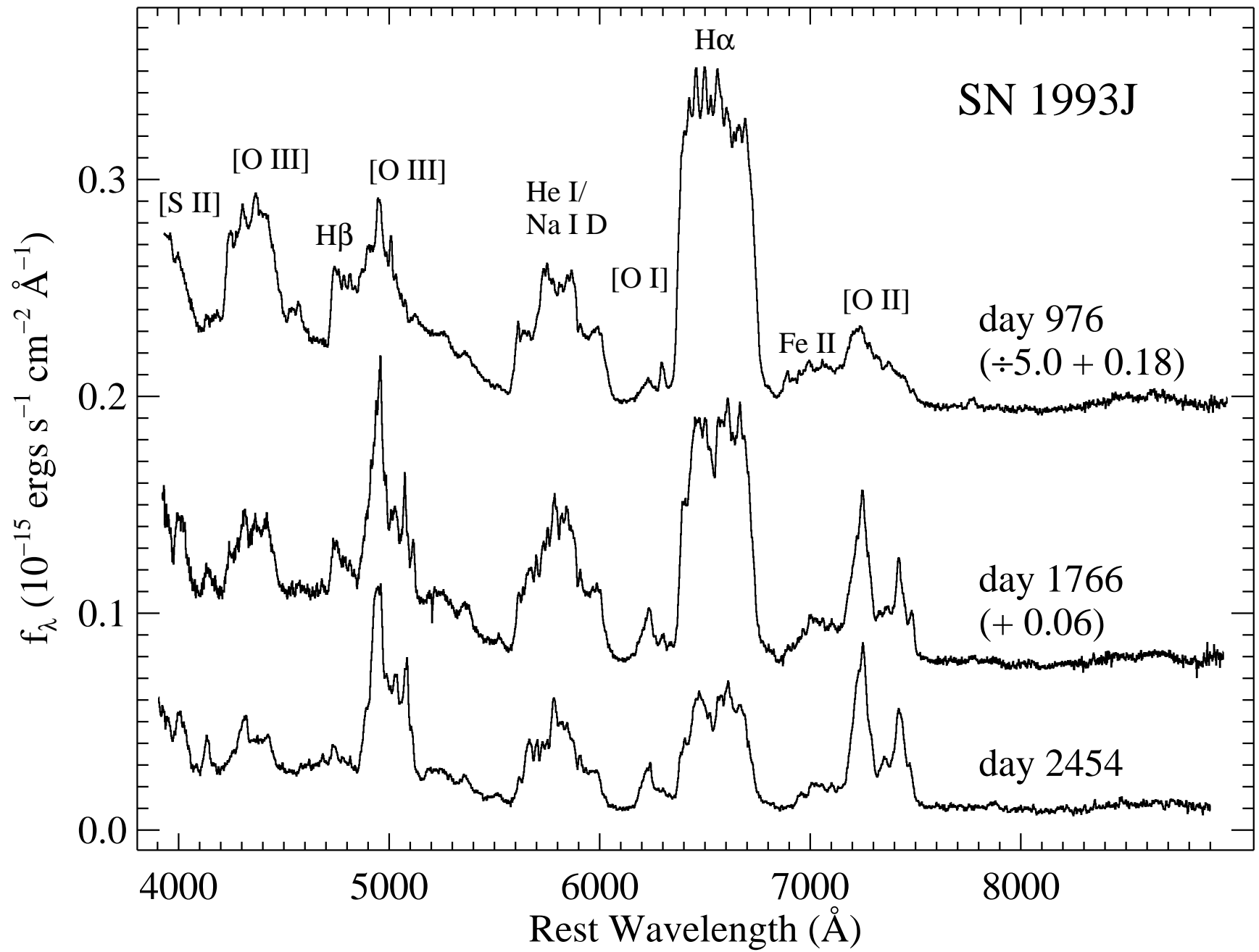
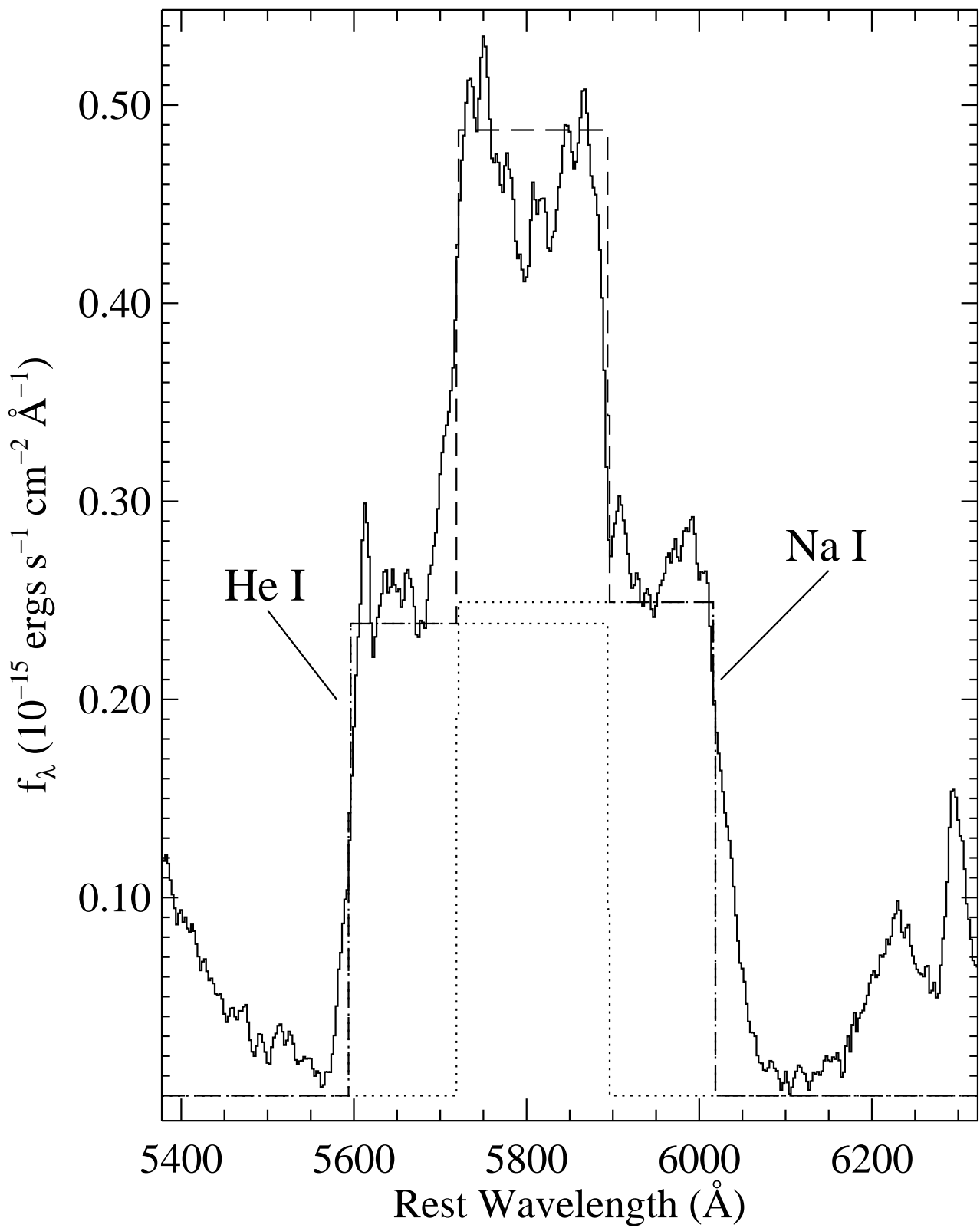
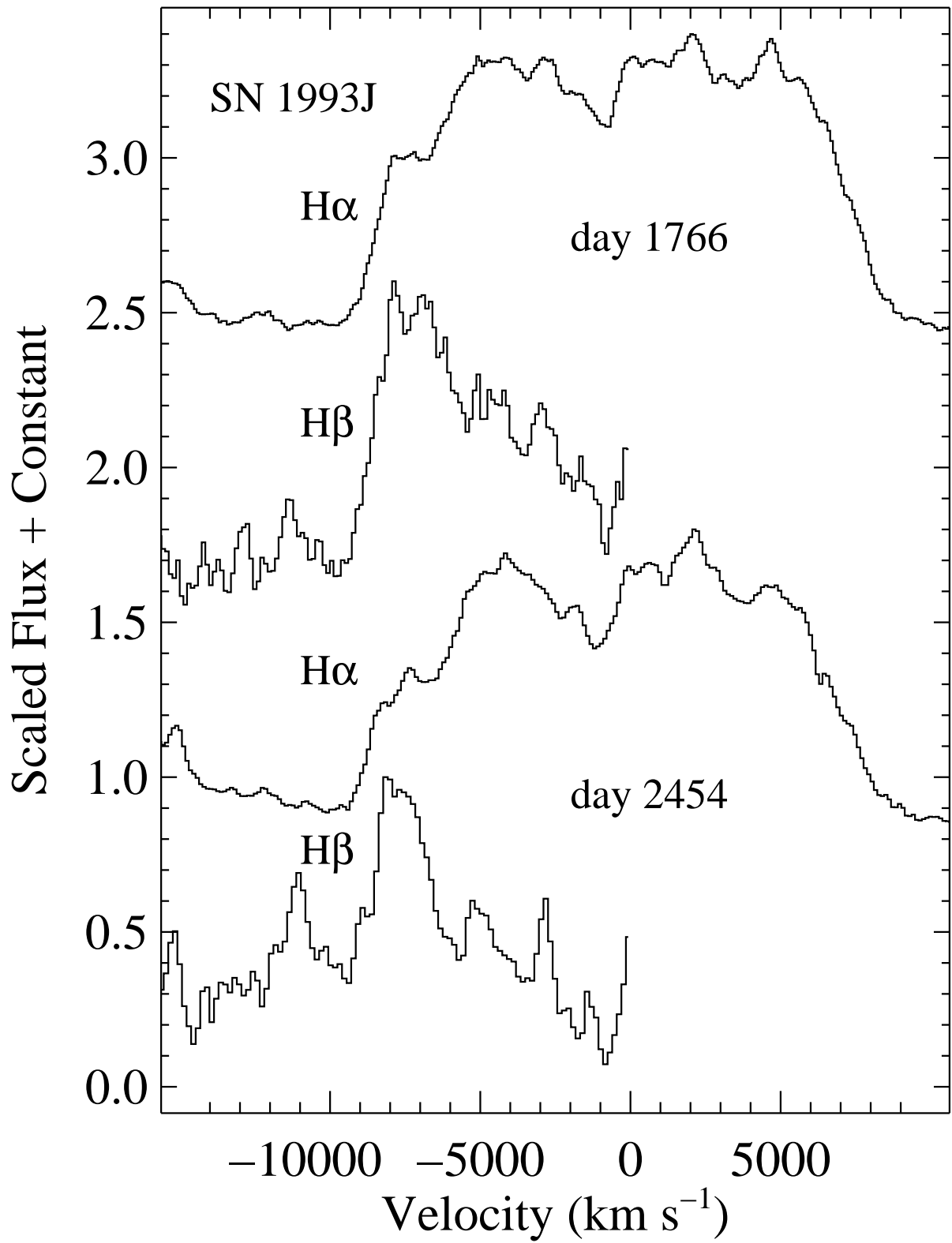
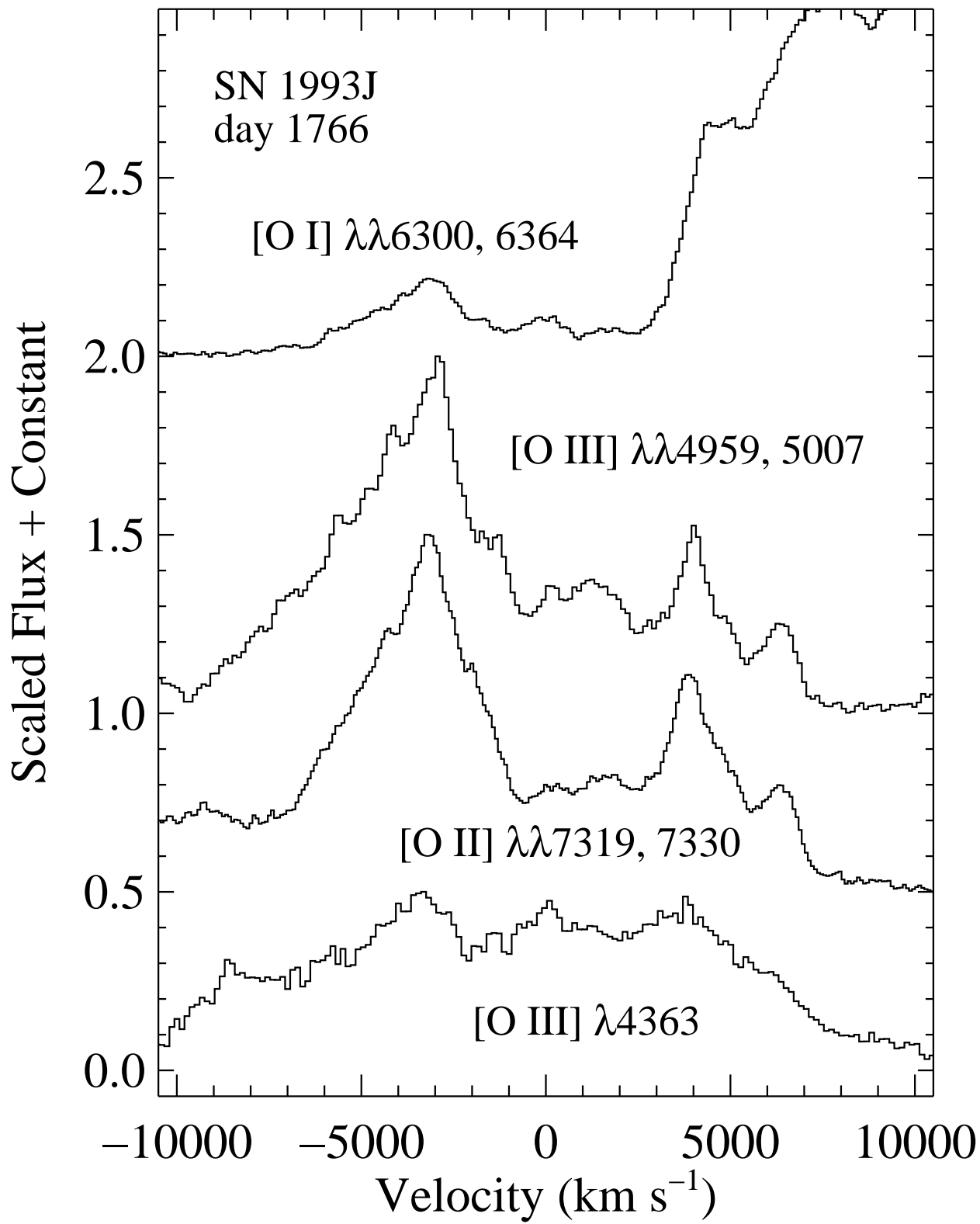
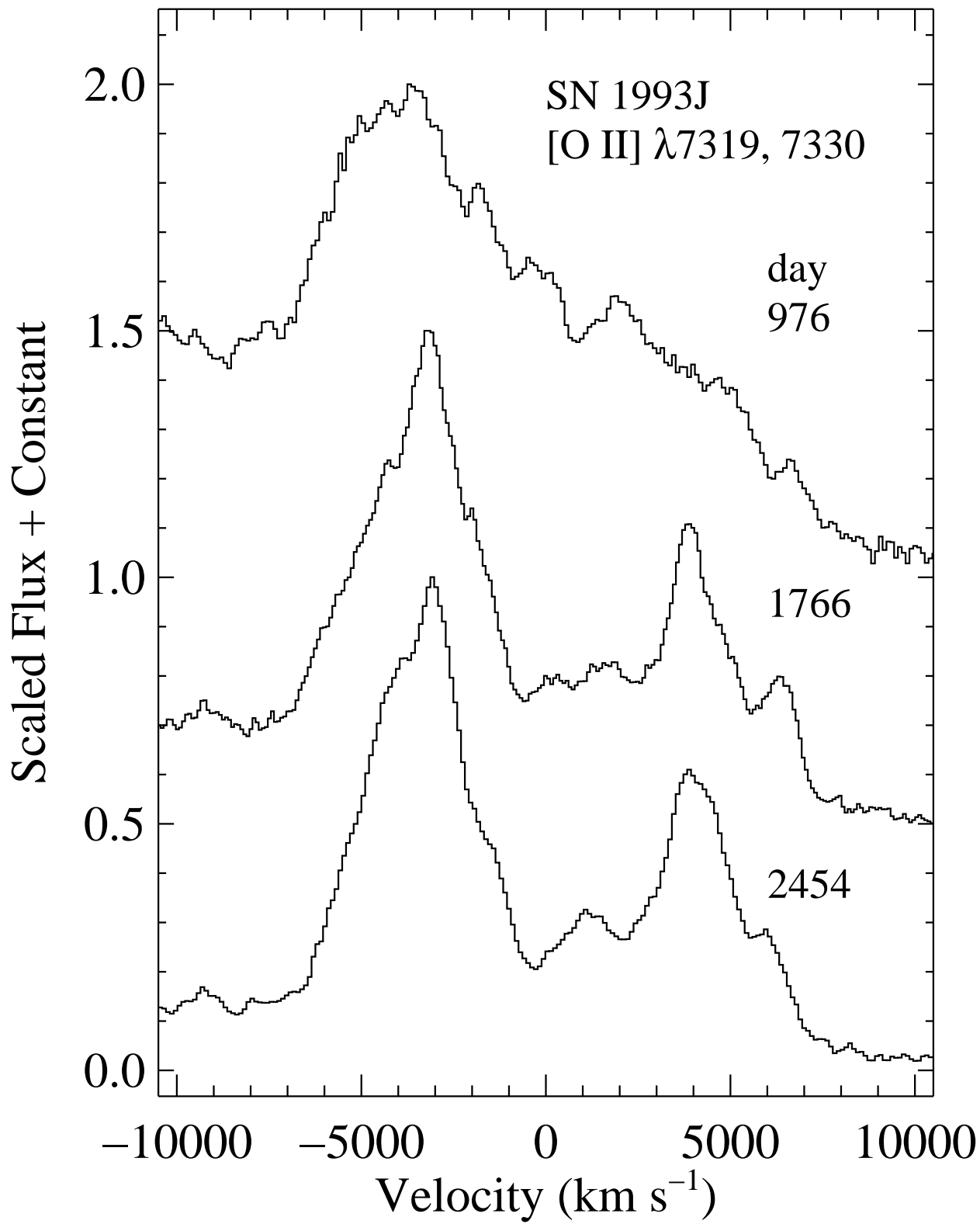
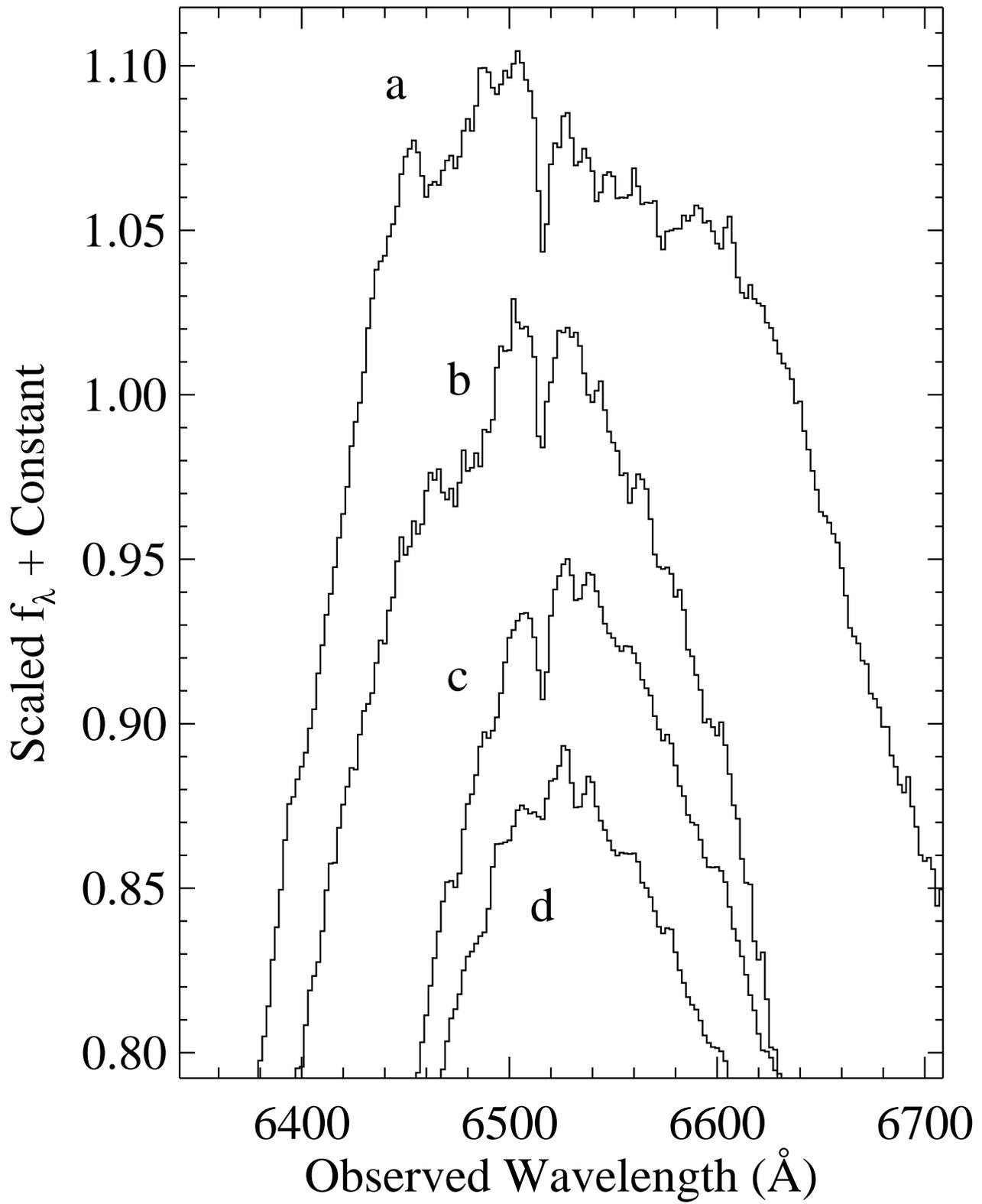
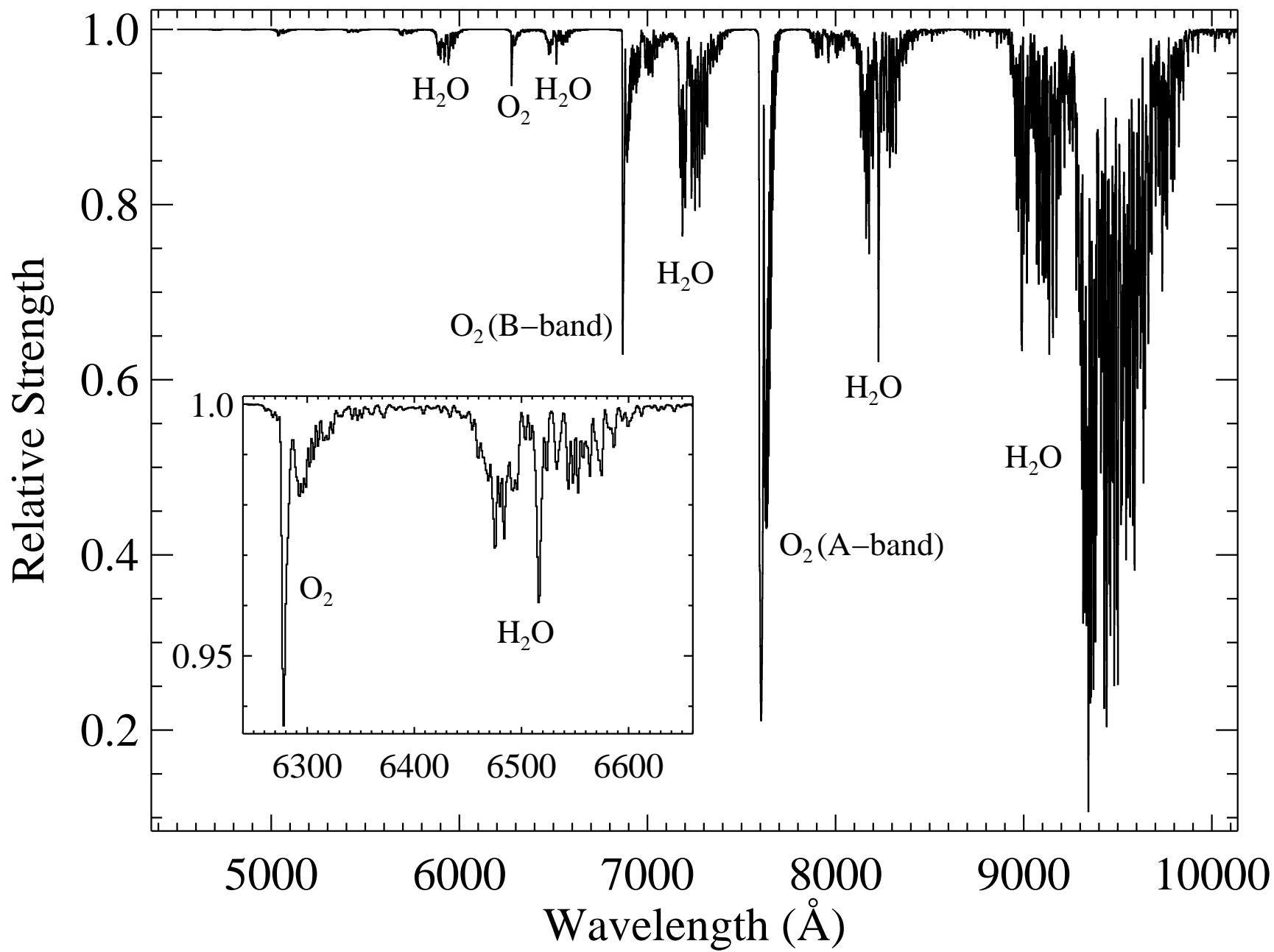
\end{document}